\documentclass[11pt,letterpaper]{article}

\usepackage[utf8]{inputenc}
\usepackage[T1]{fontenc}
\usepackage{lmodern}
\usepackage[expansion=false]{microtype}

\usepackage[margin=1in]{geometry}
\usepackage{setspace}
\usepackage{parskip}

\usepackage{amsmath,amssymb,amsthm,amsfonts}
\usepackage{mathtools}
\numberwithin{equation}{section}

\usepackage{graphicx}
\usepackage{booktabs}
\usepackage{array}
\usepackage{tabularx}
\usepackage{multirow}
\usepackage{enumitem}
\usepackage{caption}
\usepackage{float}
\usepackage{xurl}
\usepackage[table]{xcolor}
\definecolor{linkblue}{RGB}{0,68,136}
\definecolor{citeblue}{RGB}{0,90,160}
\definecolor{urlblue}{RGB}{30,80,170}
\definecolor{codekw}{RGB}{0,68,170}
\definecolor{codecm}{RGB}{120,120,120}
\definecolor{codestr}{RGB}{160,40,40}
\definecolor{codebg}{RGB}{248,248,248}
\definecolor{tabhead}{RGB}{235,240,247}
\definecolor{idgreen}{RGB}{30,132,73}
\definecolor{idamber}{RGB}{180,140,15}
\definecolor{idred}{RGB}{192,57,43}

\usepackage[colorlinks=true,
            linkcolor=linkblue,
            citecolor=citeblue,
            urlcolor=urlblue,
            breaklinks=true]{hyperref}

\usepackage{xspace}
\usepackage{xurl}

\DeclareRobustCommand{\jlcmd}[1]{\texttt{\path{#1}}\xspace}
\usepackage{listings}

\lstdefinelanguage{Julia}{%
  morekeywords={abstract,break,case,catch,const,continue,do,else,elseif,end,
  export,false,for,function,immutable,import,importall,if,in,macro,module,
  mutable,quote,return,struct,true,try,type,typealias,using,while,let,begin},
  sensitive=true,
  morecomment=[l]{\#},
  morecomment=[n]{\#=}{=\#},
  morestring=[s]{"}{"},
}

\usepackage{titlesec}
\titleformat{\section}{\large\bfseries}{\thesection.}{0.6em}{}
\titleformat{\subsection}{\normalsize\bfseries}{\thesubsection.}{0.6em}{}
\titleformat{\subsubsection}{\normalsize\bfseries\itshape}{\thesubsubsection.}{0.6em}{}

\usepackage{tikz}
\usetikzlibrary{shapes.geometric,arrows.meta,positioning,fit,backgrounds}

\newcommand{\Kid}{\mathbb{K}_{\mathrm{id}}}
\newcommand{\R}{\mathbb{R}}

\newcommand{\Glob}{\textcolor{idgreen}{\textbf{G}}}
\newcommand{\Loc}{\textcolor{idamber}{\textbf{L}}}
\newcommand{\Nid}{\textcolor{idred}{\textbf{NI}}}

\title{A Tutorial on Symbolic Structural Identifiability Analysis of ODE Models in Julia}

\author{Abdallah Alsammani\thanks{Email:
\texttt{aalsammani@desu.edu}.}\\
Department of Mathematical Sciences,
Delaware State University, Dover, DE 19901}

\date{}

\begin{document}
\maketitle


\begin{abstract}
\noindent Structural identifiability analysis determines whether the parameters of a mechanistic ordinary differential equation (ODE) model can be uniquely recovered from ideal observations and is a prerequisite for reliable parameter estimation. This tutorial presents a reproducible computational framework for symbolic structural identifiability analysis using \jlcmd{StructuralIdentifiability.jl}. We introduce the concepts of local and global identifiability, observability, parameter-to-output mappings, and identifiable parameter combinations within a unified computational workflow. The methodology is illustrated through six case studies from epidemiology, pharmacokinetics, and within-host viral dynamics, demonstrating globally identifiable models, locally but not globally identifiable models, structurally non-identifiable models, identifiable parameter combinations, and strategies for restoring identifiability through additional measurements, known initial conditions, fixed parameters, and model reparameterization. Every identifiability verdict reported here is reproduced by companion scripts run in a pinned and tested Julia environment and, where possible, confirmed by an explicit input--output or symmetry argument. The tutorial also shows how models built with ModelingToolkit.jl or Catalyst.jl, or imported from SBML, enter the analysis, how several experimental conditions can be pooled, and how structural identifiability fits into the subsequent steps of experimental design, parameter estimation, and practical identifiability analysis, providing a practical guide to structural identifiability analysis for mechanistic ODE models.
\end{abstract}


\section{Introduction}\label{sec:intro}
\subsection{Structural identifiability and mechanistic modeling}

Mechanistic ordinary differential equation (ODE) models are widely used to describe dynamical processes in epidemiology, systems biology, pharmacology and pharmacokinetics, ecology, neuroscience, and biomedical engineering. They represent the mechanisms governing biological, chemical, and physical systems through interacting state variables and parameters. After calibration to experimental or observational data, these models support forecasting, intervention analysis, uncertainty quantification, experimental design, and hypothesis testing~\cite{Walter1997,Miao2011,Raue2009}.

The reliability of these analyses depends on whether the model parameters can, in principle, be uniquely recovered from the available observations. This question is addressed by \emph{structural identifiability theory}, which studies parameter uniqueness under the assumption of ideal, noise-free, continuous measurements. A parameter is structurally identifiable if distinct parameter values cannot produce identical model outputs~\cite{BellmanAstrom1970,Ljung1994}. Otherwise, multiple parameter sets may fit the observations equally well, resulting in non-unique estimates, inflated uncertainty, ill-conditioned confidence intervals, and potentially misleading biological interpretation~\cite{Raue2009,Chis2011,Villaverde2019}.

Structural identifiability should be distinguished from \emph{practical identifiability}. The former is a property of the model equations and observation scheme, whereas the latter concerns parameter recovery from finite, noisy, and discretely sampled data. Consequently, a model may be structurally identifiable yet practically unidentifiable because of limited data quality, suboptimal experimental design, parameter correlations, or restricted observation windows~\cite{Raue2009,Hines2014,Eisenberg2014}. Structural identifiability is therefore a prerequisite for meaningful parameter estimation, since additional data cannot resolve parameters that are structurally non-identifiable under a fixed measurement scheme.

Identifiability depends not only on the model equations but also on the available measurements, known inputs, and assumptions about the initial conditions. Partial observation is the rule rather than the exception: epidemic surveillance records incidence or hospital admissions rather than the full state of a population, systems biology experiments measure a few species of a larger reaction network, and pharmacokinetic--pharmacodynamic (PK/PD) studies usually sample drug concentrations in plasma or in a single tissue (see the two-compartment model in Section~\ref{subsec:pk}). Latent states, partially observed populations, unknown initial conditions, and aggregated measurements often introduce hidden parameter symmetries and non-identifiable parameter combinations. Consequently, structural identifiability analysis is essential for model formulation, measurement selection, and experimental design~\cite{Miao2011,Tuncer2018,Villaverde2016}.

\subsection{Symbolic identifiability methods and the Julia SciML ecosystem}

Structural identifiability has been studied for more than five decades, beginning with the foundational work of Bellman and \AA str\"om~\cite{BellmanAstrom1970}, Pohjanpalo~\cite{Pohjanpalo1978}, Cobelli and DiStefano~\cite{Cobelli1980}, Walter and Pronzato~\cite{Walter1997}, Ljung and Glad~\cite{Ljung1994}, and Miao \emph{et al.}~\cite{Miao2011}. These developments have led to a wide range of symbolic methods for assessing parameter identifiability in nonlinear dynamical systems.

Most symbolic methods rest on one of two ideas, and several methods that are presented as distinct are variants of the same idea. The first expands the output in a power series around the initial time. Pohjanpalo's Taylor-series method~\cite{Pohjanpalo1978} recovers the parameters from the Taylor coefficients of the output, and these coefficients are iterated Lie derivatives of the output function along the vector field of the model. Methods phrased in terms of Lie derivatives or of the observability rank condition of Hermann and Krener~\cite{HermannKrener1977}, such as STRIKE-GOLDD~\cite{Villaverde2016}, are therefore Taylor-series methods as well. Generating-series methods, used in GenSSI~\cite{Ligon2018}, extend the same construction to models with inputs by taking Lie derivatives along the drift and the input vector fields~\cite{Chis2011}. SIAN~\cite{Hong2019SIAN,Hong2020} also works with truncated Taylor expansions, which it combines with differential algebra and a randomized algebraic test to obtain a criterion for global identifiability with a guaranteed probability of correctness. The second idea eliminates the unobserved states to obtain input--output equations, that is, relations among the outputs, the inputs, and their derivatives whose coefficients are functions of the parameters~\cite{Ljung1994}. DAISY~\cite{Bellu2007}, COMBOS~\cite{Meshkat2009,Meshkat2014}, and the projection-based elimination of Dong \emph{et al.}~\cite{Dong2023}, on which the global analysis in \jlcmd{StructuralIdentifiability.jl} is built, belong to this second family. The two families address the same question and differ in the algebraic objects they compute and in the conditions under which their conclusions hold (Section~\ref{subsec:io}).

These ideas are implemented in several established tools: DAISY in the REDUCE computer algebra system~\cite{Bellu2007}, GenSSI~2.0~\cite{Ligon2018} and STRIKE-GOLDD~\cite{Villaverde2016} as MATLAB toolboxes, COMBOS as a web application~\cite{Meshkat2014}, and SIAN in Maple, with a separate Julia implementation~\cite{Hong2019SIAN}. MATLAB and Maple are commercial systems, and, apart from the Julia version of SIAN, these tools run outside the environment in which Julia users build, simulate, and fit their models. \jlcmd{StructuralIdentifiability.jl} is an open-source Julia package that accepts models written with \jlcmd{ModelingToolkit.jl} and \jlcmd{Catalyst.jl}, so the identifiability analysis can be applied to the same model object that is later simulated and calibrated, without transcribing the equations into another system.

Symbolic structural identifiability analysis nevertheless remains underused and inconsistently applied, as documented for the infectious disease modeling literature~\cite{Liyanage2025}, in part because its results require some mathematical background to interpret and because it can become computationally expensive for large nonlinear models~\cite{Dong2023,Hong2020}.

\jlcmd{StructuralIdentifiability.jl}~\cite{Dong2023} provides algorithms for local and global identifiability analysis and for computing identifiable combinations of parameters and, more generally, observable functions of parameters and states~\cite{DeminPogudinRackauckas2026}. Models built with \jlcmd{ModelingToolkit.jl}~\cite{Ma2021MTK} or \jlcmd{Catalyst.jl}~\cite{Loman2023Catalyst} are converted to its internal representation, analyzed, and then simulated with \jlcmd{DifferentialEquations.jl}~\cite{Rackauckas2017} and calibrated with the parameter-estimation and uncertainty-quantification packages of the Julia Scientific Machine Learning (SciML) ecosystem (Section~\ref{sec:ecosystem}). Identifiability also matters when mechanistic models are combined with machine learning. Daneker \emph{et al.}~\cite{Daneker2023SBINN} pair systems-biology-informed neural networks for parameter estimation with structural and practical identifiability analysis of the underlying ODE model; Qian \emph{et al.}~\cite{Qian2024CoaguloNet} use \jlcmd{StructuralIdentifiability.jl} to determine which parameters of a blood-coagulation model a physics-informed neural network can infer from a given set of measured species; and Norden \emph{et al.}~\cite{Norden2025} show that classifiers trained on time series from partially observed dynamical systems generalize better when the parameter relations implied by structural non-identifiability are built into training. A neural network attached to a mechanistic model does not remove the structural non-identifiability of that model, so the analysis described in this tutorial remains a prerequisite in these settings.

\subsection{Objectives and contributions}

This tutorial provides a practical introduction to symbolic structural identifiability analysis using \jlcmd{StructuralIdentifiability.jl}. Rather than presenting new theoretical results, it focuses on computational methods, interpretation, and reproducible workflows for mechanistic ODE models. Specifically, the tutorial:

\begin{enumerate}[leftmargin=*,itemsep=3pt]
\item introduces the concepts of structural identifiability, observability, parameter-to-output mappings, and identifiable parameter combinations;

\item presents a unified workflow for symbolic structural identifiability analysis;

\item illustrates the methodology through six case studies from epidemiology, pharmacokinetics, and within-host viral dynamics~\cite{Alsammani2023HBV,Alsammani2025Vaccination}, together with a reaction-network model of enzyme kinetics imported from SBML;

\item discusses model reparameterization, measurement selection, and experimental design; and

\item provides reproducible Julia scripts, a notebook, and a pinned computational environment for research and graduate instruction.
\end{enumerate}

A recent tutorial by Liyanage \emph{et al.}~\cite{Liyanage2025}, published in this journal, uses the same software and shares several aims with the present one. It is organized around compartmental epidemic models (SEIR variants with asymptomatic and presymptomatic transmission, vector-borne transmission, hospitalization, and disease-induced mortality), compares unknown and known initial conditions for each model, cross-checks selected results with SIAN and STRIKE-GOLDD, and proposes compartmental diagrams annotated with identifiability results as a way to communicate them. The two tutorials overlap in their use of \jlcmd{StructuralIdentifiability.jl}, in the importance they give to initial-condition assumptions, and in treating an SEIR model with hospitalization. The present tutorial differs in four respects. First, it states the underlying mathematics explicitly (the parameter-to-output map, the field of identifiable functions, input--output equations, and the conditions under which their coefficients are identifiable) and pairs each software verdict with a symmetry or input--output argument that explains it. Second, its examples extend beyond epidemiology to pharmacokinetics, within-host viral dynamics, and a systems biology reaction network. Third, it shows how models defined with \jlcmd{ModelingToolkit.jl}, \jlcmd{Catalyst.jl}, or SBML enter the analysis, and how several experimental conditions can be pooled. Fourth, it documents a pinned and tested computational environment. Readers who work mainly with epidemic models will find in Liyanage \emph{et al.}~\cite{Liyanage2025} a larger set of worked epidemic examples and a comparison with other software for selected models.

\subsection{Organization of the paper}

The remainder of the paper is organized as follows. Section~\ref{sec:math} presents the mathematical foundations of structural identifiability. Section~\ref{sec:framework} introduces the computational workflow implemented in \jlcmd{StructuralIdentifiability.jl}, including model construction with \jlcmd{ModelingToolkit.jl}, \jlcmd{Catalyst.jl}, and SBML. Section~\ref{sec:cases} presents case studies from epidemiology, pharmacokinetics, and within-host viral dynamics. Section~\ref{sec:reparam} discusses model reparameterization and identifiable parameter combinations. Section~\ref{sec:best} summarizes best practices and limitations, Section~\ref{sec:ecosystem} describes the next steps of the Julia modeling workflow after structural identifiability analysis, and Section~\ref{sec:reproducibility} addresses reproducible computational workflows. Finally, Sections~\ref{sec:discussion} and~\ref{sec:conclusion} provide the discussion and concluding remarks.

\section{Mathematical Background}\label{sec:math}

\subsection{The class of models}\label{subsec:math-model}

We consider parametric ODE systems of the form
\begin{equation}\label{eq:ode}
\Sigma(\theta):\quad
\begin{cases}
\dot{x}(t)=f\bigl(x(t),u(t);\theta\bigr), & x(0)=x_0,\\[2pt]
y(t)=g\bigl(x(t),u(t);\theta\bigr),
\end{cases}
\qquad t\in[0,T],
\end{equation}
where $x(t)\in\R^n$ is the state vector, $u(t)\in\R^m$ is a known input, $y(t)\in\R^q$ is the measured output, and $\theta\in\Theta\subseteq\R^p$ is the parameter vector. We assume that $f$ and $g$ are rational functions of the state variables and parameters. This class includes most compartmental, mass-action, Michaelis--Menten, pharmacokinetic, within-host viral dynamics, and systems biology models. It also includes Hill-type kinetics such as $x^{n}/(K^{n}+x^{n})$ when the Hill coefficient $n$ is a fixed positive integer, because the expression is then a ratio of polynomials in $x$ and $K$. If $n$ is an unknown parameter or a non-integer number, the right-hand side is not rational and the model must first be transformed (Section~\ref{subsec:limitations}). The initial state $x_0$ is not assumed to be known; the different assumptions that can be made about it are discussed in Section~\ref{subsec:phi}.

\subsection{The parameter-to-output map}\label{subsec:phi}

For a given input $u(\cdot)$ and parameter vector $\theta$, suppose that system~\eqref{eq:ode} admits a unique solution. The corresponding output trajectory is then uniquely determined by $\theta$ and the initial condition $x_0$. We treat $x_0$ as unknown together with $\theta$. This defines the parameter-to-output map
\begin{equation}\label{eq:phi}
\Phi:\Theta\times\R^n\longrightarrow\mathcal{Y},
\qquad
(\theta,x_0)\longmapsto y(\cdot;\theta,x_0),
\end{equation}
where $\mathcal{Y}$ denotes an appropriate space of output trajectories, for example $C^\infty([0,T];\R^q)$. Structural identifiability is determined by the injectivity of $\Phi$ with respect to $\theta$. Geometrically, $\Phi$ describes the observable behavior of the model, while each fiber $\Phi^{-1}(\Phi(\theta,x_0))$ consists of pairs of parameter values and initial states that produce identical outputs and therefore cannot be distinguished from the available measurements.

Four assumptions about the initial state occur in practice, and they lead to different identifiability problems.
\begin{enumerate}[label=(\roman*),leftmargin=*,itemsep=2pt]
\item \emph{Unknown initial conditions.} The initial state is unknown and generic and is inferred together with $\theta$. This is the setting of the definitions in Section~\ref{subsec:defs}, of the theory of Hong \emph{et al.}~\cite{Hong2020}, whose definition of identifiability quantifies over generic parameter values and generic initial conditions, and of the algorithms in \jlcmd{StructuralIdentifiability.jl} and SIAN. Some presentations define identifiability with the initial conditions fixed, although the methods they describe treat the initial conditions as unknown; software verdicts should therefore be read in the augmented sense.
\item \emph{Known but generic initial conditions.} Some components of $x_0$ are measured or otherwise known, but their values are not special. This information can only enlarge the set of identifiable quantities. \jlcmd{StructuralIdentifiability.jl} provides this setting through the experimental keyword \jlcmd{known_ic} (Section~\ref{subsec:api}), which treats the declared initial values as known and generic.
\item \emph{Specific numerical initial values.} Values such as $x_2(0)=0$ for a peripheral compartment before the first dose, or $R(0)=0$ at the start of an epidemic, are known but not generic. Identifiability at such special points can differ from the generic answer, in either direction, and is covered neither by the default analysis nor by \jlcmd{known_ic}; it has to be examined for the specific values.
\item \emph{Initial conditions that depend on the parameters.} If $x_0$ is a known function of $\theta$, for example a pre-infection steady state, this relation must be imposed explicitly, because the default analysis treats $x_0$ and $\theta$ as independent.
\end{enumerate}
A verdict computed under assumption~(i) therefore holds for almost all initial states but not necessarily for a particular one, and an identifiability analysis should state which assumption was made.

\subsection{Local and global structural identifiability}\label{subsec:defs}

Structural identifiability is defined through the parameter-to-output map~\eqref{eq:phi}~\cite{Walter1997,Miao2011,Hong2020}. Here ``for almost every'' means for all points outside a set of Lebesgue measure zero; for the rational models considered here, the exceptional set can be taken to lie in a proper algebraic subset, which is how Hong \emph{et al.}~\cite{Hong2020} state the definitions. Let $h(\theta)$ denote a parameter $\theta_i$ or, more generally, a rational function of the parameters.

\begin{description}[leftmargin=*,style=nextline]
\item[Globally identifiable.] The quantity $h$ is \emph{globally structurally identifiable} if, for almost every $(\theta,x_0)\in\Theta\times\R^n$,
\begin{equation}\label{eq:globalid}
\Phi(\tilde{\theta},\tilde{x}_0)=\Phi(\theta,x_0)
\;\Longrightarrow\;
h(\tilde{\theta})=h(\theta)
\qquad
\text{for all }(\tilde{\theta},\tilde{x}_0)\in\Theta\times\R^n.
\end{equation}

\item[Locally identifiable.] The quantity $h$ is \emph{locally structurally identifiable} if, for almost every $(\theta,x_0)\in\Theta\times\R^n$, there exists a neighborhood $U$ of $(\theta,x_0)$ such that the implication in~\eqref{eq:globalid} holds for all $(\tilde{\theta},\tilde{x}_0)\in U$. Hong \emph{et al.}~\cite{Hong2020} use the equivalent formulation that, for almost every $(\theta,x_0)$, the set $\{h(\tilde{\theta}) : \Phi(\tilde{\theta},\tilde{x}_0)=\Phi(\theta,x_0)\}$ is finite.

\item[Non-identifiable.] The quantity $h$ is \emph{structurally non-identifiable} if it is not locally identifiable. For almost every $(\theta,x_0)$, every neighborhood of $(\theta,x_0)$ then contains points that produce the same output but a different value of $h$, so the values of $h$ compatible with the output form a continuum.
\end{description}

A model is globally (respectively locally) identifiable if every parameter is globally (respectively locally) identifiable. Local identifiability is characterized by the full rank of an appropriate sensitivity matrix~\cite{Sedoglavic2002} and is a generic algebraic property. This criterion is the observability rank condition of Hermann and Krener~\cite{HermannKrener1977} applied to the model extended by the parameters as additional states with zero derivative: a parameter is locally identifiable when the Jacobian, with respect to the states and parameters, of the outputs and their successive Lie derivatives loses rank once that parameter is removed from the differentiation variables. Sedoglavic's algorithm evaluates this rank at a random point using truncated power series (Section~\ref{subsec:io}). Global identifiability additionally requires the absence of discrete parameter symmetries, such as parameter exchange in compartmental models or sign symmetries in models with squared terms.

Intuitively, local identifiability guarantees a unique solution only within a neighborhood of the true parameter values, whereas global identifiability guarantees uniqueness throughout the parameter space. Consequently, locally but not globally identifiable parameters can be recovered only up to a finite ambiguity.

\subsection{Observability and the augmented formulation}\label{subsec:augmented}

When the initial conditions are unknown, both the parameter vector $\theta$ and the initial state $x_0$ must be inferred from the observed outputs. This naturally introduces the concept of \emph{observability}. A state variable $x_i(t)$ is observable if it can be uniquely determined, for almost every $(\theta,x_0)$, from the measured outputs and known inputs.

Because the definitions of Section~\ref{subsec:defs} are stated for the augmented unknown $(\theta,x_0)\in\Theta\times\R^n$, they apply in the same way to the initial values: $x_i$ is globally or locally observable when $x_i(0)$ satisfies the corresponding definition, and the whole trajectory $x_i(t)$ is then determined by the output up to the same ambiguity. \jlcmd{StructuralIdentifiability.jl} reports the observability of states in the same format as the identifiability of parameters.

Known initial conditions provide additional algebraic constraints that often restore identifiability. Consequently, whether an initial condition is declared known or unknown can substantially affect the analysis, as the viral dynamics example of Section~\ref{subsec:viral} shows.

\subsection{Identifiable functions and reparameterization}\label{subsec:identifiable-functions}

Structural non-identifiability does not imply that a model is uninformative. Even when individual parameters cannot be uniquely recovered, certain functions of the parameters may still be uniquely determined by the model output. Formally, a rational function $\phi$ of the parameters is identifiable if it is globally identifiable in the sense of~\eqref{eq:globalid} with $h=\phi$.

The identifiable rational functions form a subfield
\[
\mathcal{K}_{\mathrm{id}}\subseteq\mathbb{Q}(\theta).
\]
Since every subfield of the finitely generated field $\mathbb{Q}(\theta)$ is itself finitely generated, $\mathcal{K}_{\mathrm{id}}$ admits a finite set of generators. This is a general fact about subfields of a field of rational functions and does not depend on the identifiability problem. What depends on the model is how generators can be found. The coefficients of suitably normalized input--output equations (Section~\ref{subsec:io}) generate the field of functions that are identifiable from several independent experiments~\cite{Ovchinnikov2022}, and this field coincides with $\Kid$ under a condition discussed in Section~\ref{subsec:io}. \jlcmd{StructuralIdentifiability.jl} computes these coefficients by the differential elimination of Dong \emph{et al.}~\cite{Dong2023} and then replaces them by a simpler generating set of the same field using the algorithm of Demin and Pogudin~\cite{DeminPogudin2025,DeminPogudin2026}.

Functions of the parameters and the states can be treated in the same way. A rational function $h(\theta,x)$ is observable if its value $h(\theta,x(t))$ along the solution is determined by the output, in the sense of~\eqref{eq:globalid} with $h(\theta,x(t))$ in place of $h(\theta)$; for a function of $\theta$ alone this is identifiability as defined above, and for a single state it is observability (Section~\ref{subsec:augmented}). These functions form a field, called the observation field by Demin, Pogudin, and Rackauckas~\cite{DeminPogudinRackauckas2026}, which contains $\Kid$. The two fields answer different questions: $\Kid$ describes which parameter combinations can be estimated, and the observation field also describes which combinations of the states, such as a rescaled population, can be reconstructed from the data. Demin, Pogudin, and Rackauckas prove that generators of the observation field can be obtained from the identifiable functions of the parameters together with the coefficients of Lie derivatives of the outputs, and their algorithm is implemented in \jlcmd{StructuralIdentifiability.jl} (Section~\ref{subsec:api}).

These combinations naturally define a reparameterization of the model. Replacing the original parameters with
\[
\eta=(\eta_1,\ldots,\eta_r), \qquad \eta_j=\phi_j(\theta),
\]
produces an equivalent parameterization in which the identifiable quantities are explicit. A classical example, revisited in Section~\ref{sec:reparam}, is the model
\[
\dot{x}=-pq\,x,
\]
where only the product $pq$ is identifiable.

\subsection{Differential algebra and input-output equations}\label{subsec:io}

Global structural identifiability of nonlinear ODE models is commonly analyzed using differential algebra~\cite{Ljung1994,Bellu2007}. The central idea is to eliminate the unobserved state variables by repeatedly differentiating the output equations, yielding an input-output equation of the form
\begin{equation}\label{eq:io}
P\bigl(y,\dot y,\ddot y,\ldots,u,\dot u,\ldots;c(\theta)\bigr)=0,
\end{equation}
where $P$ is a polynomial whose coefficients $c(\theta)$ are rational functions of the parameters, with one coefficient normalized to one.

Input--output equations are central to identifiability analysis because every identifiable function of the parameters can be written in terms of their coefficients: identifiability implies input--output identifiability~\cite{OvchinnikovPogudinThompson2021}. The converse, however, is not automatic, and the coefficients $c(\theta)$ are not always identifiable. They are always identifiable from several independent experiments with generic initial conditions, since they generate the field of multi-experiment identifiable functions~\cite{Ovchinnikov2022}, but they need not be identifiable from a single experiment. Example~2.14 of Hong \emph{et al.}~\cite{Hong2020} shows this with a model of one state,
\[
\dot x=0,\qquad y_1=x,\qquad y_2=\theta x+\theta^2 .
\]
Its input--output equations $\dot y_1=0$ and $y_2-\theta y_1-\theta^2=0$ contain $\theta$ as a coefficient, yet a single experiment produces only the constants $y_1=x(0)$ and $y_2=\theta x(0)+\theta^2$, and $\theta$ is then one of the two roots $\theta$ and $-\theta-x(0)$ of a quadratic equation. The parameter is locally but not globally identifiable, although its value would follow uniquely from the coefficients. Single-experiment and input--output identifiability coincide when the model has no rational first integrals~\cite{OvchinnikovPogudinThompson2021} (in Example~2.14 the state $x$ is itself a first integral), and a sufficient condition that can be checked algorithmically is the nonsingularity of a Wronskian matrix built from the monomials of the input--output equations~\cite{OvchinnikovPogudinThompson2021,Dong2023}. In the benchmark collection of Dong \emph{et al.}~\cite{Dong2023}, this check succeeded for every model but one (their Table~3), so the exception is uncommon in applications but not merely hypothetical.

The older statement that the coefficients of an input--output equation are always identifiable is therefore too strong. Software that relies on it without checking the condition can report global identifiability where only local identifiability holds; Hong \emph{et al.}~\cite{Hong2020} show that DAISY does so for Example~2.14. \jlcmd{StructuralIdentifiability.jl} computes input--output equations by a probabilistic differential elimination algorithm based on projections~\cite{Dong2023}, performs the Wronskian check, and prints a warning when the check fails; its global verdicts then refer to multi-experiment identifiability. Deciding whether a given parameter, or a given function of the parameters, belongs to the field generated by $c(\theta)$ is a field-membership problem, which the package solves with randomized algorithms whose probability of error is controlled by the user~\cite{Dong2023}. Running Example~2.14 through the package shows why the warning matters:
\begin{lstlisting}
ode = @ODEmodel(
    x'(t)  = 0,
    y1(t)  = x(t),
    y2(t)  = a * x(t) + a^2
)
assess_identifiability(ode)
# Warning: One of the Wronskians has corank greater than one, so the
#   results of the algorithm will be valid only for multiexperiment
#   identifiability. [...]
# x(t) => :globally
# a    => :globally
\end{lstlisting}
The verdict for $a$ is correct for two experiments with different initial values, which together remove the two-fold ambiguity, but a single experiment identifies $a$ only locally. When the warning appears, the reported global verdicts should therefore be read in the multi-experiment sense, and single-experiment identifiability should be checked by other means; the warning itself suggests SIAN or rewriting the parameters as states with zero derivative.

Local structural identifiability is assessed using the probabilistic algorithm of Sedoglavic~\cite{Sedoglavic2002}. The method computes truncated power-series solutions of the model and of its sensitivity equations at randomly chosen parameter values and initial conditions and checks the rank of the resulting Jacobian modulo a large prime. Its cost grows polynomially with the size of the model, which makes it the natural first step; for large models it can nevertheless take considerable time (Section~\ref{subsec:limitations}).

\subsection{Summary of the analysis workflow}

The structural identifiability workflow used throughout this tutorial consists of the following steps:

\begin{enumerate}[leftmargin=*,itemsep=2pt]
\item \textbf{Specify the model and measurements.} Define the ODE system~\eqref{eq:ode}, including the measured outputs, known inputs, and any initial conditions assumed to be known.

\item \textbf{Assess local identifiability.} Apply the Sedoglavic test. Parameters that are locally non-identifiable are structurally non-identifiable, so they need no global test as individual parameters. Functions that involve them may still be identifiable and are found in step~4.

\item \textbf{Assess global identifiability.} For locally identifiable parameters, apply the differential algebra-based test to detect discrete parameter ambiguities. In \jlcmd{StructuralIdentifiability.jl}, \jlcmd{assess_identifiability} runs the local test internally before the global one.

\item \textbf{Compute identifiable functions.} If some parameter is not globally identifiable, compute the generators of $\Kid$ to identify suitable parameter combinations for reparameterization.

\item \textbf{Interpret the results.} Use the identifiability analysis to guide model reparameterization, measurement selection, parameter fixing, or experimental design.
\end{enumerate}

This workflow is implemented in \jlcmd{StructuralIdentifiability.jl} and is followed throughout the case studies in Section~\ref{sec:cases}. A graphical overview is provided in Figure~\ref{fig:workflow}.

\section{Computational Framework}\label{sec:framework}

\subsection{The Julia ecosystem for scientific computing}

Julia~\cite{Bezanson2017} is a high-level, just-in-time compiled programming language designed for scientific computing. Several features make it well suited for structural identifiability analysis. It generates efficient native code through LLVM while maintaining a high-level programming interface. Its multiple-dispatch paradigm and expressive type system support generic algorithms that operate on floating-point, automatic differentiation, and symbolic data types. In addition, the Julia Scientific Machine Learning (SciML) ecosystem provides a unified environment for symbolic computation, ODE solving, parameter estimation, uncertainty quantification, and identifiability analysis~\cite{Rackauckas2017}.

Within the Julia SciML ecosystem, the interoperability of \jlcmd{StructuralIdentifiability.jl} runs in one direction. Models written with the symbolic modeling package \jlcmd{ModelingToolkit.jl}~\cite{Ma2021MTK} or the reaction-network package \jlcmd{Catalyst.jl}~\cite{Loman2023Catalyst} are converted into the internal representation of \jlcmd{StructuralIdentifiability.jl} and analyzed there (Section~\ref{subsec:mtk}). The converse does not hold: a model created with \jlcmd{@ODEmodel} is a data structure of \jlcmd{StructuralIdentifiability.jl}, not a ModelingToolkit model, and cannot be passed to \jlcmd{DifferentialEquations.jl}~\cite{Rackauckas2017} for simulation or to estimation packages. The package exports such models only to the input formats of other identifiability tools (Maple/SIAN, DAISY, COMBOS, and GenSSI). In a typical workflow the model is therefore written once in ModelingToolkit or Catalyst, analyzed with \jlcmd{StructuralIdentifiability.jl}, and then simulated and calibrated in its original form (Section~\ref{sec:ecosystem}).

\subsection{\texttt{StructuralIdentifiability.jl}}

\jlcmd{StructuralIdentifiability.jl}~\cite{Dong2023,Ovchinnikov2022} is a Julia package for symbolic structural identifiability analysis of rational ODE models. It provides the following functionality:

\begin{itemize}[leftmargin=*,itemsep=2pt]
\item \textbf{Local identifiability analysis} using the probabilistic polynomial-time algorithm of Sedoglavic \cite{Sedoglavic2002}.

\item \textbf{Global identifiability analysis} using a probabilistic differential elimination algorithm with user-specified correctness probability~\cite{Dong2023}, followed by the Wronskian check of Section~\ref{subsec:io}.

\item \textbf{Analysis of user-defined rational functions} of the model parameters.

\item \textbf{Computation of generators of identifiable functions} for the model parameters and, optionally, the state variables.

\item \textbf{Interfaces to model-building packages:} ModelingToolkit \jlcmd{System} objects through a package extension of \jlcmd{StructuralIdentifiability.jl}, and Catalyst \jlcmd{ReactionSystem} objects through an extension of \jlcmd{Catalyst.jl}.
\end{itemize}

The package supports rational ODE models with multiple inputs and multiple outputs. Initial conditions are unknown by default, and an experimental option declares selected initial conditions known and generic. All analyses reported in this tutorial were performed using Julia~1.12.2 and \jlcmd{StructuralIdentifiability.jl} v0.5.26. The complete computational environment, including the exact versions of all dependencies, is recorded in the \texttt{Project.toml} and \texttt{Manifest.toml} files of the companion scripts and notebook and archived in the reproducibility package (Section~\ref{sec:reproducibility}). The underlying algorithms are randomized, and the probability that a result is incorrect is bounded by a user-specified threshold. \jlcmd{find_identifiable_functions} seeds its random number generator (Section~\ref{subsec:api}), whereas \jlcmd{assess_identifiability} and \jlcmd{assess_local_identifiability} draw from Julia's global random number generator, which the companion scripts seed explicitly. Runtimes vary across systems and executions.

\subsection{Core API}\label{subsec:api}

The principal functions used throughout this tutorial are summarized below.

\subsubsection{\texttt{@ODEmodel}}

The \jlcmd{@ODEmodel} macro provides a concise mathematical syntax for defining ODE models:

\begin{lstlisting}
using StructuralIdentifiability

ode = @ODEmodel(
    x1'(t) = -k * x1(t),
    x2'(t) =  k * x1(t) - g * x2(t),
    y(t)   =  x2(t)
)
\end{lstlisting}

Differentiated variables define the model states, variables declared as \jlcmd{u(t)} are treated as inputs, and all remaining symbols are interpreted as parameters. Output equations of the form \jlcmd{y(t)=...} specify the measured quantities. Multiple inputs and outputs are defined by adding additional equations.

\paragraph{Multiple observables.}

Additional outputs often improve structural identifiability by providing independent algebraic information. Multiple observables are declared by adding additional output equations:

\begin{lstlisting}
ode = @ODEmodel(
    x1'(t) = -k1 * x1(t) + k2 * x2(t),
    x2'(t) =  k1 * x1(t) - k2 * x2(t),
    y1(t)  =  x1(t),
    y2(t)  =  x2(t)
)
\end{lstlisting}

Internally, all output equations contribute to the input-output elimination procedure, so the global analysis uses the combined information from all measured outputs. Additional observables can resolve parameter ambiguities that are present under a single measurement, as illustrated by the viral dynamics, SIWR, and SEIR-H case studies (Sections~\ref{subsec:viral}, \ref{subsec:siwr}, and~\ref{subsec:seirh}).

\subsubsection{\texttt{assess\_local\_identifiability}}

The function \jlcmd{assess_local_identifiability(ode)} implements Sedoglavic's probabilistic algorithm~\cite{Sedoglavic2002}. It returns a dictionary that maps each parameter and each state to \jlcmd{true} (locally identifiable or observable) or \jlcmd{false} (not identifiable even locally). The keyword argument \jlcmd{prob_threshold} specifies the desired correctness probability (default: 0.99), and \jlcmd{loglevel} controls the amount of printed diagnostics. The keyword \jlcmd{type} selects single-experiment (\jlcmd{:SE}, the default) or multi-experiment (\jlcmd{:ME}) local identifiability; the latter refers to repeated experiments with the same parameters and independent initial conditions and is defined for parameters only. The cost of the test grows polynomially with the size of the model~\cite{Sedoglavic2002}, so it is usually much cheaper than the global analysis and is the recommended first step; Section~\ref{subsec:limitations} discusses what this means for large models.

\subsubsection{\texttt{assess\_identifiability}}

The function \jlcmd{assess_identifiability(ode)} determines whether each parameter or initial state is \jlcmd{:globally}, \jlcmd{:locally}, or \jlcmd{:nonidentifiable}. Specific rational functions can be analyzed using the keyword argument \jlcmd{funcs_to_check}, for example,

\begin{lstlisting}
funcs_to_check = [a01 + a12, a01 * a12]
\end{lstlisting}

Global analysis is performed using the probabilistic differential elimination algorithm described by Dong \emph{et al.}~\cite{Dong2023}; the correctness probability is again controlled by \jlcmd{prob_threshold}. The function first runs the local test and applies the global test only to the quantities found to be locally identifiable. If the Wronskian check of Section~\ref{subsec:io} fails, it prints a warning that the result is valid only for multi-experiment identifiability. An experimental keyword argument \jlcmd{known_ic} allows selected initial conditions to be declared known and generic (assumption~(ii) of Section~\ref{subsec:phi}); the states passed to it must belong to the same model object. This is the mechanism used in Section~\ref{subsec:viral} to remove a scaling ambiguity.

\subsubsection{\texttt{find\_identifiable\_functions}}

The function \jlcmd{find_identifiable_functions(ode)} computes a generating set of $\Kid$. These generators determine every identifiable rational function of the model parameters. For the parameters, the computation follows the procedure described in Section~7.1 of Demin and Pogudin~\cite{DeminPogudin2026} and has two stages. First, input--output equations are computed by differential elimination~\cite{Dong2023}, and their normalized coefficients are taken as an initial generating set; these coefficients generate the field of multi-experiment identifiable functions~\cite{Ovchinnikov2022}, which equals $\Kid$ when the Wronskian check of Section~\ref{subsec:io} succeeds. Second, this initial set, which is often large and hard to interpret, is replaced by a simpler set of generators of the same field. The simplification algorithm of Demin and Pogudin~\cite{DeminPogudin2025,DeminPogudin2026} computes only the low-degree part of a Gr\"obner basis of an associated polynomial ideal, using evaluation and sparse interpolation, and adds candidate generators found by searching the field for polynomials of small degree; it improves on the earlier approach of Ovchinnikov \emph{et al.}~\cite{Ovchinnikov2022}, which computes the complete set of Gr\"obner-basis coefficients and then removes redundant ones. The keyword \jlcmd{simplify} (\jlcmd{:weak}, \jlcmd{:standard}, \jlcmd{:strong}, or \jlcmd{:absent}) trades the simplicity of the result against computing time. Setting \jlcmd{with_states=true} extends the computation to observable functions of both parameters and state variables, that is, to generators of the observation field of Section~\ref{subsec:identifiable-functions}. The package then implements the algorithm of Demin, Pogudin, and Rackauckas~\cite{DeminPogudinRackauckas2026}: it starts from the identifiable functions of the parameters, adds the coefficients of Lie derivatives of the outputs up to orders read off from the input--output equations, and simplifies the result. The case studies report parameter combinations, because these are what parameter estimation and reparameterization require; Section~\ref{subsec:viral} shows the state functions for one model.

A generating set is not unique: any set of functions that generates the same field is equally valid, so the generators reported in this tutorial are one convenient representation, and a different package version or option may return a different set that generates the same field. The algorithms are Monte Carlo algorithms whose probability of returning an incorrect result is bounded through \jlcmd{prob_threshold}, but the function seeds its random number generator (keyword \jlcmd{seed}, default 42), so repeated calls with the same model, options, and package version return the same set. In our checks with the single-output models of Sections~\ref{subsec:pk}--\ref{subsec:seirh}, five repeated calls, five other seeds, and the three simplification levels other than \jlcmd{:absent} all returned the generating sets reported in Section~\ref{sec:cases}.

Conceptually, \jlcmd{assess_identifiability} determines which parameters are identifiable, whereas \jlcmd{find_identifiable_functions} identifies the parameter combinations that can be estimated.

\subsection{Building models with \texttt{ModelingToolkit.jl}, \texttt{Catalyst.jl}, and SBML}\label{subsec:mtk}

Although \jlcmd{@ODEmodel} provides the most direct way to define models for structural identifiability analysis, many Julia users build models with \jlcmd{ModelingToolkit.jl}~\cite{Ma2021MTK} or, for reaction networks, \jlcmd{Catalyst.jl}~\cite{Loman2023Catalyst}. Models of both kinds can be passed to \jlcmd{assess_identifiability}, \jlcmd{assess_local_identifiability}, and \jlcmd{find_identifiable_functions}, which convert them to the internal representation of \jlcmd{StructuralIdentifiability.jl}; the conversion runs in this direction only (Section~\ref{sec:framework}). The outputs shown below were obtained with \jlcmd{ModelingToolkit.jl} 11.38.1, \jlcmd{Catalyst.jl} 16.4.0, and \jlcmd{SBMLImporter.jl} 4.1.2 (Section~\ref{sec:reproducibility}).

\paragraph{ModelingToolkit.} A ModelingToolkit \jlcmd{System} (called \jlcmd{ODESystem} in earlier releases) is passed together with the \jlcmd{measured_quantities} keyword, which lists the observed quantities as symbolic equations. For the SIR model of Section~\ref{subsec:sir}:
\begin{lstlisting}
using ModelingToolkit, StructuralIdentifiability

@independent_variables t
D = Differential(t)
@parameters beta gamma
@variables S(t) I(t) R(t) y(t)
eqs = [D(S) ~ -beta * S * I,
       D(I) ~  beta * S * I - gamma * I,
       D(R) ~  gamma * I]
@named sir_mtk = System(eqs, t)

assess_identifiability(sir_mtk; measured_quantities = [y ~ beta * S * I])
# S(t) => :globally, I(t) => :globally, R(t) => :nonidentifiable,
# beta => :globally, gamma => :globally
\end{lstlisting}
The call \jlcmd{mtk_to_si(sir_mtk, [y ~ beta * S * I])} performs the conversion explicitly and returns the \jlcmd{StructuralIdentifiability.jl} model together with a dictionary that relates the variables of the two representations, which is useful when lower-level functions of the package are needed.

\paragraph{Catalyst.} Catalyst \jlcmd{ReactionSystem} objects are supported through a package extension that is loaded automatically when both packages are imported. The measured quantities are given as species (for example \jlcmd{:I} or \jlcmd{sir_rn.I}) or as expressions of species and parameters:
\begin{lstlisting}
using Catalyst, StructuralIdentifiability

sir_rn = @reaction_network begin
    beta,  S + I --> 2I
    gamma, I --> R
end

assess_identifiability(sir_rn;
    measured_quantities = [sir_rn.beta * sir_rn.S * sir_rn.I])
# S(t) => :globally, I(t) => :globally, R(t) => :nonidentifiable,
# beta => :globally, gamma => :globally
\end{lstlisting}
Because the reaction network generates exactly the ODE system~\eqref{eq:sir} with the same measured output, the parameter verdicts coincide with those of the \jlcmd{@ODEmodel} implementation in Section~\ref{subsec:sir}. Two details of the extension are worth knowing. By default it eliminates conservation laws before the analysis (here $S+I+R$ is constant, and the converted model returned by \jlcmd{make_si_ode} contains equations for $S$ and $I$ only) and then reports the results in terms of the original species; the option \jlcmd{remove_conserved = false} skips the elimination. Parameters whose values are known can be declared with the keyword \jlcmd{known_p}.

\paragraph{SBML.} The same workflow also supports standardized model exchange. \jlcmd{SBMLImporter.jl}~\cite{Bostrom2024SBMLImporter} imports SBML models as Catalyst \jlcmd{ReactionSystem} objects, allowing models distributed in the SBML format to be analyzed without manual reimplementation. The reproducibility package contains an SBML file of the Michaelis--Menten mechanism $E+S\rightleftharpoons C\rightarrow E+P$ with mass-action kinetics:
\begin{lstlisting}
using SBMLImporter, Catalyst, StructuralIdentifiability

rn, callbacks = load_SBML("models/enzyme_kinetics.xml")
assess_identifiability(rn; measured_quantities = [:P],
    known_p = [:cell], remove_conserved = false)
# S(t), C(t), P(t), E(t) => :globally
# k1, km1, k2, cell      => :globally
\end{lstlisting}
With the product concentration measured and all initial concentrations unknown, the three rate constants and the four concentrations are globally identifiable. Two practical points arose in this pipeline. The compartment volume \jlcmd{cell} is imported as a model parameter and has to be declared known through \jlcmd{known_p}; otherwise it would be treated as an unknown. In the versions used here, \jlcmd{SBMLImporter.jl} imports stoichiometric coefficients symbolically, and Catalyst's conservation-law elimination does not accept symbolic stoichiometry, so the analysis is run with \jlcmd{remove_conserved = false}. SBML models may also contain events, piecewise definitions, or algebraic constraints; these are not part of the rational ODE class of Section~\ref{subsec:math-model}, and the Catalyst extension currently rejects reaction systems coupled with algebraic equations. Reaction networks stored in the BioNetGen \texttt{.net} format or as stoichiometric matrices can be loaded with \jlcmd{ReactionNetworkImporters.jl}~\cite{ReactionNetworkImporters}, which also returns a Catalyst \jlcmd{ReactionSystem} but does not read SBML files. Section~\ref{sec:ecosystem} describes how this workflow extends to simulation, sensitivity analysis, parameter estimation, and Bayesian inference.

\subsection{Computational workflow}\label{subsec:pipeline}

\begin{figure}[!ht]
  \centering
  \includegraphics[width=\linewidth]{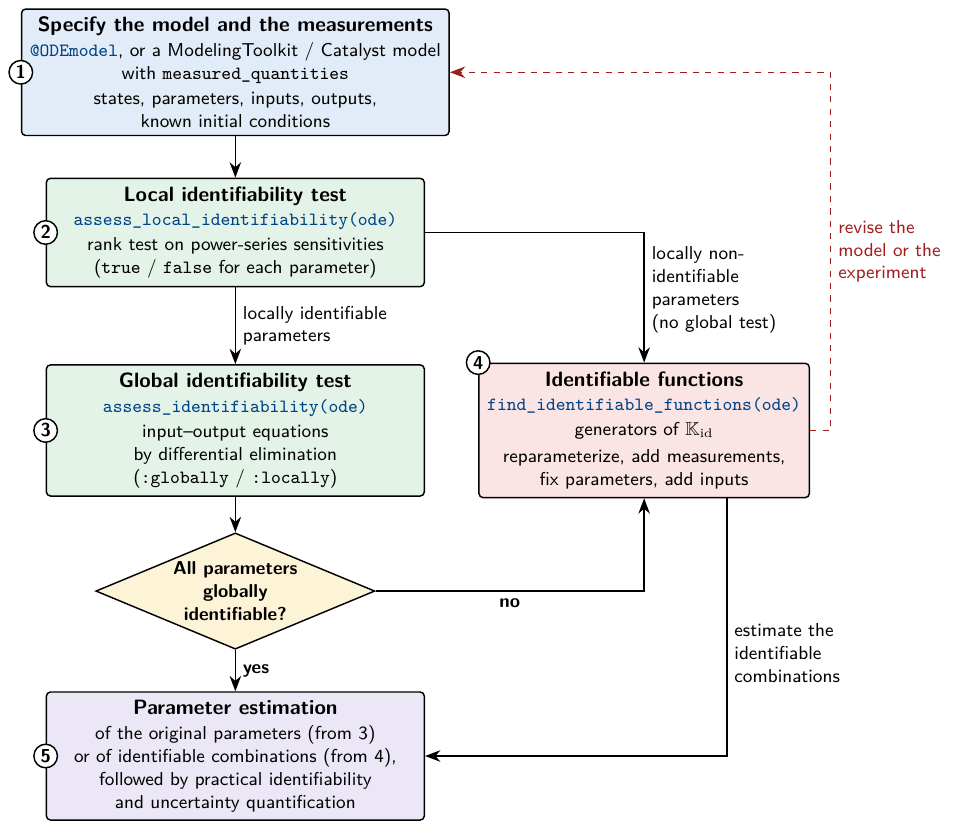}
  \caption{\textbf{Structural identifiability workflow implemented in \jlcmd{StructuralIdentifiability.jl}.} Starting from a model defined with \jlcmd{@ODEmodel} (or converted from a ModelingToolkit or Catalyst model), local identifiability is assessed using Sedoglavic's algorithm. Parameters that are not even locally identifiable need no global test and lead directly to the identifiable functions; the locally identifiable parameters are tested for global identifiability using probabilistic differential elimination. Parameters that are only locally identifiable also lead to the identifiable functions, which suggest a reparameterization, additional measurements or inputs, or parameters to fix (dashed arrow), or which can be estimated in place of the original parameters. Parameter estimation is therefore applied to the original parameters when all of them are globally identifiable, and to identifiable combinations otherwise.}
  \label{fig:workflow}
\end{figure}

The accompanying notebook implements the computational workflow summarized in Table~\ref{tab:pipeline} and applies it consistently to all case studies. Figure~\ref{fig:workflow} provides a graphical overview of the analysis pipeline.

\begin{table}[!ht]
\centering
\caption{Recommended workflow for structural identifiability analysis using \jlcmd{StructuralIdentifiability.jl}. The same procedure is applied throughout the case studies in Section~\ref{sec:cases}.}
\label{tab:pipeline}
\renewcommand{\arraystretch}{1.30}
\setlength{\tabcolsep}{6pt}
\small
\begin{tabularx}{\linewidth}{@{}c l X@{}}
\toprule
\rowcolor{tabhead}
\textbf{Step} & \textbf{Julia call} & \textbf{Purpose} \\
\midrule

1 &
\texttt{ode = @ODEmodel(...)} &
Define the model, measured outputs, inputs, and known initial conditions, or pass a ModelingToolkit or Catalyst model with \texttt{measured\_quantities} (Section~\ref{subsec:mtk}). \\

2 &
\jlcmd{assess_local_identifiability(ode)} &
Assess local identifiability using Sedoglavic's algorithm. Parameters that are not locally identifiable need no global test; combinations involving them are found in step~4. \\

3 &
\jlcmd{assess_identifiability(ode)} &
Assess global identifiability and detect discrete parameter ambiguities. \\

4 &
\jlcmd{find_identifiable_functions(ode)} &
Compute identifiable parameter combinations for model reparameterization when needed. \\

5 &
(optional) &
Repeat the analysis with selected initial conditions declared known, if appropriate. \\

6 &
(interpret) &
Use the results to guide measurement selection, parameter fixing, reparameterization, or experimental design. \\

\bottomrule
\end{tabularx}
\end{table}


\section{Case Studies}\label{sec:cases}

\begin{figure}[!ht]
  \centering
  \includegraphics[width=\linewidth]{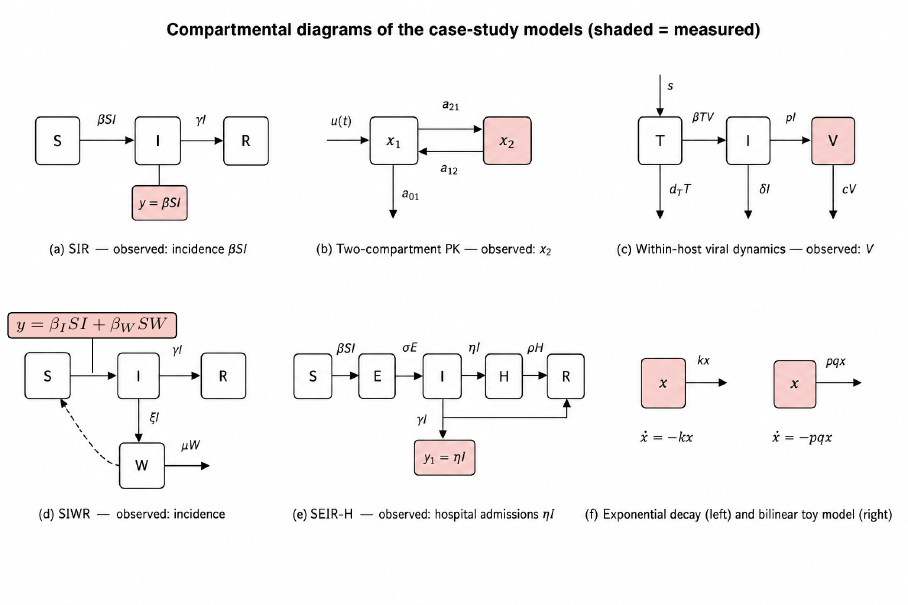}
  \caption{\textbf{Compartmental diagrams of the models analyzed in Section~\ref{sec:cases}.} Shaded compartments or shaded output boxes are measured; arrows are labeled by the corresponding flows. (a) SIR model with incidence $\beta SI$ observed; $\beta$ and $\gamma$ are globally identifiable. (b) Two-compartment pharmacokinetic model with a known input $u(t)$ into the central compartment and the peripheral compartment $x_2$ observed; the exchange symmetry $a_{01}\leftrightarrow a_{12}$ makes these two parameters locally but not globally identifiable, while $a_{21}$ is globally identifiable. (c) Within-host viral dynamics model with viral load $V$ observed; $\beta$ and $d_T$ are globally identifiable, $c$ and $\delta$ are locally identifiable because they can be exchanged, and $s$ and $p$ enter only through the product $ps$. (d) SIWR model with aggregated incidence $y=\beta_I SI+\beta_W SW$ observed; $\beta_I$ is globally identifiable, $\gamma$ and $\mu$ can be exchanged, and $\beta_W$ and $\xi$ enter only through the combination $\beta_I\mu+\beta_W\xi$. (e) SEIR-H model with hospital admissions $\eta I$ observed; a scaling of $(S,E,I)$ leaves the output invariant, so $\beta$, $\eta$, and $\gamma$ are non-identifiable, and $\rho$ cannot be identified without observing $H$. (f) Exponential decay (globally identifiable) and the bilinear toy model of Section~\ref{sec:reparam} (only $pq$ identifiable).}
  \label{fig:compartments}
\end{figure}

The case studies are arranged in approximately increasing complexity. Each illustrates a distinct aspect of structural identifiability: a globally identifiable benchmark, global identifiability from a nonlinear output, local but not global identifiability caused by an exchange symmetry, continuous non-identifiability caused by a scaling symmetry of unobserved states, identifiable combinations that mix parameters of different transmission pathways, and the coexistence of discrete and continuous ambiguities in a model used for epidemic forecasting. Figure~\ref{fig:compartments} summarizes the models. For every model, the initial conditions are treated as unknown unless stated otherwise, which is the default assumption of \jlcmd{StructuralIdentifiability.jl}; several examples show that declaring an initial condition known changes the verdict.

Each case study follows a common structure comprising the scientific background, model formulation, Julia implementation, identifiability results, a short explanation of \emph{why} the verdict holds, and a discussion of the implications for parameter estimation and experimental design. The explanations are deliberately elementary: a symmetry that leaves the output invariant proves non-identifiability, and an input--output equation shows which combinations of parameters the output determines; that these combinations are identifiable from a single experiment is what the Wronskian check of Section~\ref{subsec:io} confirms, and it succeeded for every model in this section. Presenting both the software output and the underlying argument guards against misreading a probabilistic verdict and is the practice we recommend. A consolidated summary is provided in Table~\ref{tab:cases} and Figure~\ref{fig:verdicts}.

\subsection{Exponential decay: a minimal benchmark}\label{subsec:expdecay}

\paragraph{Background.}
The exponential decay model is the simplest ODE encountered in pharmacokinetics, radioactive decay, first-order chemical kinetics, and many related applications. It serves as a benchmark because its parameter is globally identifiable.

\paragraph{Model.}
Let $x(t)$ denote the state variable and $k>0$ the decay rate:
\begin{equation}\label{eq:expdecay}
\dot{x}(t)=-k\,x(t),\qquad
x(0)=x_0,\qquad
y(t)=x(t).
\end{equation}

\paragraph{Julia implementation.}\, \\
\begin{lstlisting}
ode = @ODEmodel(
    x'(t) = -k * x(t),
    y(t)  =  x(t)
)
\end{lstlisting}

\paragraph{Results.} Both \jlcmd{assess_local_identifiability} and \jlcmd{assess_identifiability} classify both the parameter $k$ and the state $x(t)$ as globally identifiable, returning \jlcmd{k => :globally} and \jlcmd{x(t) => :globally}. This result follows directly from the analytical solution $x(t)=x_0e^{-kt}$. When the initial condition $x_0$ is known, the decay rate $k$ is uniquely determined by the output trajectory. If $x_0$ is unknown but treated as an additional parameter, continuous observation of $x(t)$, or equivalently measurements at two distinct time points, uniquely determines both $x_0$ and $k$. Consequently, the model is globally structurally identifiable.

\paragraph{Discussion.}
This example serves as a basic validation of the analysis workflow. Any structural identifiability method should classify the model as globally identifiable, and \jlcmd{StructuralIdentifiability.jl} does so immediately. More importantly, the example illustrates that structural identifiability is determined by the parameter-to-output map rather than by the amount of available data.

\subsection{The SIR model: a classical compartmental example}\label{subsec:sir}

\paragraph{Background.}
The Susceptible--Infectious--Recovered (SIR) model is the classical compartmental model of infectious disease transmission~\cite{Kermack1927}. It describes epidemic dynamics through two parameters: the transmission rate $\beta$ and the recovery rate $\gamma$. In practice, the most commonly observed quantity is the incidence of new infections,
\[
\beta S I.
\]

\paragraph{Model.}
With state variables $(S,I,R)$ and constant total population, the model is
\begin{equation}\label{eq:sir}
\begin{aligned}
\dot S &= -\beta SI,\\
\dot I &= \beta SI-\gamma I,\\
\dot R &= \gamma I,
\end{aligned}
\end{equation}
with observed output
\begin{equation}\label{eq:sir-out}
y(t)=\beta S(t)I(t).
\end{equation}

\paragraph{Julia implementation.}\, \\
\begin{lstlisting}
ode = @ODEmodel(
    S'(t) = -beta * S(t) * I(t),
    I'(t) =  beta * S(t) * I(t) - gamma * I(t),
    R'(t) =  gamma * I(t),
    y(t)  =  beta * S(t) * I(t)
)
\end{lstlisting}

\paragraph{Results.}
The function \jlcmd{assess_identifiability} classifies both $\beta$ and $\gamma$ as globally identifiable, while correctly reporting $R$ as unobservable because it does not appear in the input-output equations generated from $y$. Consequently, the basic reproduction number,
\[
R_0=\frac{\beta S_0}{\gamma},
\]
is also structurally identifiable when $S_0$ is known.

\paragraph{Discussion.}
This example illustrates an important feature of structural identifiability. Although the recovered compartment is unobserved, incidence data alone are sufficient to identify both transmission and recovery rates. This follows from the input-output equations obtained by eliminating the unobserved states, which depend only on $\beta$, $\gamma$, the observed output, and its derivatives. These equations are generated automatically by \jlcmd{StructuralIdentifiability.jl}~\cite{Dong2023}. As shown in Section~\ref{subsec:mtk}, the same conclusions are obtained when the model is defined as a \jlcmd{Catalyst.jl} reaction network.

\subsection{A two-compartment pharmacokinetic model}\label{subsec:pk}

\paragraph{Background.}
Linear compartmental models are widely used in pharmacokinetics to describe drug distribution between the bloodstream and peripheral tissues. The two-compartment model is the classical benchmark for structural identifiability because, for a suitable measurement scheme, it exhibits local but not global identifiability owing to a parameter exchange symmetry~\cite{BellmanAstrom1970,Cobelli1980,Walter1997}. The model analyzed here is a two-state version of the introductory example in the \jlcmd{StructuralIdentifiability.jl} README, which adds a third state.

\paragraph{Model.}
Let $x_1$ denote the central compartment, into which a known input $u(t)$ (for example, an infusion rate in amount per unit time) is administered, and $x_2$ a peripheral compartment whose concentration is measured, as in tissue sampling or microdialysis studies:
\begin{equation}\label{eq:pk}
\begin{aligned}
\dot{x}_1 &= -(a_{01}+a_{21})x_1+a_{12}x_2+u,\\
\dot{x}_2 &= a_{21}x_1-a_{12}x_2,\\
y &= x_2,
\end{aligned}
\end{equation}
where $a_{21}$ and $a_{12}$ are the intercompartmental transfer rates and $a_{01}$ is the elimination rate. Both initial conditions are unknown.

\paragraph{Julia implementation.}\, \\
\begin{lstlisting}
ode = @ODEmodel(
    x1'(t) = -(a01 + a21) * x1(t) + a12 * x2(t) + u(t),
    x2'(t) =  a21 * x1(t) - a12 * x2(t),
    y(t)   =  x2(t)
)
\end{lstlisting}

\paragraph{Results.}
The function \jlcmd{assess_identifiability(ode)} returns

\begin{lstlisting}
x1(t)  => :locally
x2(t)  => :globally
a01    => :locally
a12    => :locally
a21    => :globally
\end{lstlisting}

The parameters $a_{01}$ and $a_{12}$ are locally but not globally identifiable because the output is invariant under the exchange $a_{01}\leftrightarrow a_{12}$ (with $x_1(0)$ adjusted accordingly). Only their symmetric combinations,
\[
a_{01}+a_{12} \quad\text{and}\quad a_{01}a_{12},
\]
are globally identifiable, as confirmed by

\begin{lstlisting}
assess_identifiability(
    ode,
    funcs_to_check = [a01 + a12, a01 * a12]
)
\end{lstlisting}

and \jlcmd{find_identifiable_functions(ode)} returns the generating set $\{a_{21},\,a_{01}+a_{12},\,a_{01}a_{12}\}$~\cite{Ovchinnikov2022,Meshkat2009}.

\paragraph{Why.}
Eliminating $x_1$ from~\eqref{eq:pk} gives the input--output equation
\[
\ddot y+(a_{01}+a_{12}+a_{21})\,\dot y+a_{01}a_{12}\,y=a_{21}\,u .
\]
For a generic input, the three coefficients can be recovered from a single output trajectory (the Wronskian check of Section~\ref{subsec:io} succeeds), so $a_{21}$, $a_{01}+a_{12}$, and $a_{01}a_{12}$ are globally identifiable. The individual rates $a_{01}$ and $a_{12}$ are the two roots of $z^2-(a_{01}+a_{12})z+a_{01}a_{12}=0$ and can therefore be recovered only up to their order: a two-fold ambiguity, which is exactly local identifiability. The state $x_1=(\dot y+a_{12}y)/a_{21}$ inherits the two-fold ambiguity of $a_{12}$.

\paragraph{Measurement scheme matters.}
If the central compartment is observed instead, $y=x_1$, and the input enters with an unknown gain $b$, the input--output equation becomes
\[
\ddot y+(a_{01}+a_{12}+a_{21})\,\dot y+a_{01}a_{12}\,y=b\,\dot u+b\,a_{12}\,u,
\]
whose four coefficients determine $b$, $a_{12}$, $a_{01}$, and $a_{21}$ uniquely. Accordingly, \jlcmd{assess_identifiability} classifies every parameter and both states of this variant as globally identifiable, and \jlcmd{find_identifiable_functions} returns the parameters themselves; the companion notebook contains this variant. Without the input, on the other hand, observing either compartment from unknown initial conditions identifies only $a_{01}+a_{12}+a_{21}$ and $a_{01}a_{12}$, so all three rates are structurally non-identifiable. The same dynamics thus range from globally identifiable to non-identifiable depending on the measurement and input design.

\paragraph{Discussion.}
This example illustrates that a locally identifiable model can still support meaningful inference through globally identifiable combinations. Rather than estimating $a_{01}$ and $a_{12}$ individually, one can estimate $(a_{21},\,a_{01}+a_{12},\,a_{01}a_{12})$, or resolve the ambiguity by fixing one of the two rates from independent experimental data. Otherwise, optimization may converge to either of two pharmacologically distinct parameter sets that produce identical peripheral concentration profiles, as illustrated in Figure~\ref{fig:nonidentifiability}(B).

\subsection{Within-host viral dynamics}\label{subsec:viral}

\paragraph{Background.}
The within-host viral dynamics model introduced by Perelson and colleagues~\cite{Perelson1996} is one of the most influential models in mathematical biology. It describes interactions among uninfected target cells, infected cells, and free virus, and has been widely applied to HIV, hepatitis B and C, SARS-CoV-2, and other viral infections. Structural identifiability of this model has been studied extensively because only a subset of the state variables is typically observed~\cite{Miao2011}.

\paragraph{Model.}
A standard formulation is
\begin{equation}\label{eq:viral}
\begin{aligned}
\dot T &= s-d_TT-\beta TV,\\
\dot I &= \beta TV-\delta I,\\
\dot V &= pI-cV,
\end{aligned}
\end{equation}
where viral load is the only measured quantity,
\begin{equation}\label{eq:viral-out}
y(t)=V(t),
\end{equation}
and all initial conditions are unknown.

\paragraph{Julia implementation.}\, \\
\begin{lstlisting}
ode = @ODEmodel(
    T'(t) = s - d_T * T(t) - beta * T(t) * V(t),
    I'(t) = beta * T(t) * V(t) - delta * I(t),
    V'(t) = p * I(t) - c * V(t),
    y(t)  = V(t)
)
\end{lstlisting}

\paragraph{Results.}
With viral load as the only measurement, \jlcmd{assess_identifiability} returns
\begin{lstlisting}
T(t)  => :nonidentifiable
I(t)  => :nonidentifiable
V(t)  => :globally
beta  => :globally
c     => :locally
d_T   => :globally
delta => :locally
p     => :nonidentifiable
s     => :nonidentifiable
\end{lstlisting}
and \jlcmd{find_identifiable_functions(ode)} returns the generating set
\[
\{\,d_T,\;\beta,\;c+\delta,\;c\delta,\;ps\,\}.
\]
Two qualitatively different ambiguities coexist. The clearance and death rates $c$ and $\delta$ are locally but not globally identifiable: only their sum and product are determined, so the two values can be exchanged. The production rate $p$ and the source rate $s$ are structurally non-identifiable, and so are the unobserved states $T$ and $I$; only the product $ps$ is identifiable.

\paragraph{Why.}
The continuous ambiguity is a scaling symmetry of the unobserved states. For any $\lambda>0$, the transformation
\[
(T,\,I,\,s,\,p)\;\longmapsto\;(\lambda T,\,\lambda I,\,\lambda s,\,p/\lambda),
\]
with $\beta$, $d_T$, $\delta$, and $c$ unchanged, maps solutions of~\eqref{eq:viral} to solutions with the same $V(t)$: every term of the $T$ and $I$ equations is multiplied by $\lambda$, and $pI$ is unchanged. The output therefore cannot fix the scale of $T$ and $I$, and $s$ and $p$ are determined only through $ps$. The discrete ambiguity follows from the input--output equation obtained by eliminating $I$,
\[
\ddot V+(c+\delta)\dot V+c\delta\,V=p\beta\,TV,
\]
in which $c$ and $\delta$ appear only symmetrically. Note that $\beta$ is globally identifiable even though $p$ is not: the coupling $\beta TV$ is not rescaled by the symmetry. For the model as written, with unknown initial conditions, the identifiable product is therefore $ps$ rather than $\beta p$; which product appears depends on how the model is normalized, which is one more reason to compute rather than guess.

The same two ambiguities determine which functions of the states can be recovered. With \jlcmd{with_states=true} (Section~\ref{subsec:api}), the package returns generators of the observation field:
\begin{lstlisting}
find_identifiable_functions(ode; with_states = true)
# d_T, beta, V(t), c + delta, p*s, c*delta, T(t)*p, I(t)*p + V(t)*delta
\end{lstlisting}
The products $pT$ and $ps$ are invariant under the scaling symmetry, so the target-cell population is determined only through $pT$, that is, up to the unknown viral production rate $p$. The function $pI+\delta V$ equals $\dot V+(c+\delta)V$ by the equation for $V$ and is therefore determined by the output, whereas $pI=\dot V+cV$ is determined only up to the exchange of $c$ and $\delta$.

\paragraph{Restoring identifiability.}
Suppose measurements of the target-cell population are also available,
\[
y_2(t)=T(t),
\]
as in studies that combine viral load with flow cytometry. The corresponding model is

\begin{lstlisting}
ode_augmented = @ODEmodel(
    T'(t) = s - d_T * T(t) - beta * T(t) * V(t),
    I'(t) = beta * T(t) * V(t) - delta * I(t),
    V'(t) = p * I(t) - c * V(t),
    y1(t) = V(t),
    y2(t) = T(t)
)
\end{lstlisting}

and \jlcmd{assess_identifiability(ode_augmented)} returns
\begin{lstlisting}
T(t)  => :globally
I(t)  => :locally
V(t)  => :globally
beta  => :globally
c     => :locally
d_T   => :globally
delta => :locally
p     => :globally
s     => :globally
\end{lstlisting}
Comparing this output with the single-output result above, $s$, $d_T$, $\beta$, and $p$ are globally identifiable and $T$ and $V$ are observable, because $\dot T=s-d_TT-\beta TV$ is now an equation in measured quantities and $p\beta$ is determined by the input--output equation above. The exchange symmetry of $c$ and $\delta$, however, survives: both remain locally identifiable, and the infected-cell trajectory $I$ is observable only up to the same two-fold ambiguity. Resolving it requires information that distinguishes viral clearance from infected-cell death, for example an independent estimate of $c$ from perturbation experiments~\cite{Perelson1996}, or a measurement of infected cells. Because the ambiguity has only two branches, which exchange the values of $c$ and $\delta$, knowing which of the two rates is larger is already enough to select one branch. For HIV-1, Perelson \emph{et al.}~\cite{Perelson1996} estimated a mean life-span of about 0.3~days for plasma virions and 2.2~days for productively infected cells, which supports $c>\delta$. Such an ordering is external information: it is supplied by the modeler, not inferred from the viral-load data, and the selected branch is only as reliable as the evidence for the ordering. Alternatively, the scaling ambiguity alone can be removed without a second observable by declaring $T(0)$ known, for instance from a pre-treatment cell count. In \jlcmd{StructuralIdentifiability.jl} this is expressed with the \jlcmd{known_ic} keyword argument, applied to the single-output model:
\begin{lstlisting}
T0 = only(filter(v -> string(v) == "T(t)", ode.x_vars))  # the state T of `ode`
assess_identifiability(ode; known_ic = [T0])
# T(0) => :globally,  I(0) => :locally,  V(0) => :globally
# beta, d_T, p, s => :globally;  c, delta => :locally
\end{lstlisting}
The continuous ambiguity disappears, and $s$ and $p$ become globally identifiable, while the $c\leftrightarrow\delta$ exchange persists. The known value is treated as generic (assumption~(ii) of Section~\ref{subsec:phi}). The state passed to \jlcmd{known_ic} must belong to the same model object; the first line retrieves it from \jlcmd{ode}, because after the definition of \jlcmd{ode_augmented} the name \jlcmd{T} refers to a state of that model.

\paragraph{Discussion.}
This example demonstrates that structural identifiability depends on the measurement scheme as much as on the model itself, and that a single model can harbor both a continuous and a discrete ambiguity. Viral load alone identifies $\beta$, $d_T$, $c+\delta$, $c\delta$, and $ps$; measuring target cells separates $s$ from $p$; and separating $c$ from $\delta$ needs either prior knowledge or an additional observable. Reporting this structure explicitly is more informative than reporting a single ``non-identifiable'' label~\cite{Wieland2021}.

\subsection{Environmentally transmitted disease: an SIWR model}\label{subsec:siwr}

\paragraph{Background.}
For environmentally transmitted pathogens, including \emph{Vibrio cholerae}, rotavirus, norovirus, and hepatitis A virus, the classical SIR model is extended with an environmental reservoir $W$. The resulting SIWR model is widely used to study waterborne disease transmission and has become a standard example in structural identifiability because the environmental reservoir introduces additional parameter dependencies~\cite{Eisenberg2013}.

\paragraph{Model.}
The SIWR model is
\begin{equation}\label{eq:siwr}
\begin{aligned}
\dot S &= -\beta_I SI-\beta_W SW,\\
\dot I &= \beta_I SI+\beta_W SW-\gamma I,\\
\dot W &= \xi I-\mu W,\\
\dot R &= \gamma I,
\end{aligned}
\end{equation}
where $\beta_I$ is the direct transmission rate, $\beta_W$ is the environmental transmission rate, $\xi$ is the pathogen shedding rate, $\mu$ is the environmental decay rate, and $\gamma$ is the recovery rate. The observed output is the aggregated incidence
\begin{equation}\label{eq:siwr-out}
y(t)=\beta_I S(t)I(t)+\beta_W S(t)W(t).
\end{equation}

\paragraph{Julia implementation.}\, \\
\begin{lstlisting}
ode = @ODEmodel(
    S'(t) = -beta_I*S(t)*I(t) - beta_W*S(t)*W(t),
    I'(t) =  beta_I*S(t)*I(t) + beta_W*S(t)*W(t)
             - gamma*I(t),
    W'(t) =  xi*I(t) - mu*W(t),
    R'(t) =  gamma*I(t),
    y(t)  =  beta_I*S(t)*I(t) + beta_W*S(t)*W(t)
)
\end{lstlisting}

\paragraph{Results.}
When only aggregated incidence is observed, \jlcmd{assess_identifiability} classifies $\beta_I$ as globally identifiable, $\gamma$ and $\mu$ as locally but not globally identifiable, and $\beta_W$ and $\xi$ (together with the states $W$ and $R$) as non-identifiable. The generating set returned by \jlcmd{find_identifiable_functions} is
\[
\{\,\beta_I,\;\gamma+\mu,\;\gamma\mu,\;\beta_I\mu+\beta_W\xi\,\}.
\]
Two mechanisms are at work. First, the reservoir concentration is measured in no particular unit: rescaling $(W,\beta_W,\xi)\mapsto(W/\lambda,\lambda\beta_W,\xi/\lambda)$ leaves the incidence unchanged, so $\beta_W$ and $\xi$ are individually non-identifiable. Second, the recovery rate $\gamma$ and the environmental decay rate $\mu$ can be exchanged: the subsystem formed by $I$ and $W$ is a linear two-compartment chain driven by incidence, and only the sum and product of its two rates are determined. This discrete ambiguity is not a plain exchange of two parameters. Writing $\kappa=\beta_W\xi$ for the combined strength of environmental transmission, the generators are preserved by
\[
(\gamma,\,\mu,\,\kappa)\;\longmapsto\;\bigl(\mu,\,\gamma,\,\kappa+\beta_I(\mu-\gamma)\bigr),
\]
so exchanging the recovery and decay rates must be compensated by a change in environmental transmission and in the initial state, where $I(0)$ is replaced by $I(0)+(\gamma-\mu)\beta_WW(0)/\kappa$. Together with the rescaling of $W$, the parameter values compatible with incidence data therefore form two disjoint one-dimensional curves related by this transformation. Figure~\ref{fig:nonidentifiability}(D) checks the construction by simulation: with $\beta_I=0.3$, the parameter sets $(\gamma,\mu,\beta_W\xi)=(0.5,0.25,0.2)$ and $(0.25,0.5,0.125)$, started from matched initial states, produce incidence curves that differ by less than $10^{-13}$, although their numbers of infected individuals differ by up to $0.09$ (as a fraction of the population). The last generator couples the two transmission pathways and is neither a simple product nor a simple sum: it states that incidence data cannot distinguish a pathogen with weak environmental transmission and high shedding from one with strong environmental transmission and low shedding~\cite{Eisenberg2013}. This mixed generator is the kind of identifiable combination that is difficult to guess by inspection and that \jlcmd{find_identifiable_functions} returns automatically.

\paragraph{Additional measurements.}
If environmental pathogen concentrations are also measured,
\[
y_2(t)=W(t),
\]
as in water testing or wastewater surveillance, the unit of $W$ is fixed: \jlcmd{assess_identifiability} then classifies $\xi$ as globally identifiable and $W$ as observable. The exchange of $\gamma$ and $\mu$, however, is not resolved by this measurement, and because $\beta_W$ is recovered from $\beta_I\mu+\beta_W\xi$ it inherits the same two-fold ambiguity; $\beta_W$, $\gamma$, and $\mu$ remain locally identifiable. The reason is visible in the input--output relation of the $I$--$W$ chain, $\ddot W+(\gamma+\mu)\dot W+\gamma\mu\,W=\xi\,y$, which is symmetric in $\gamma$ and $\mu$. Observing prevalence $I(t)$ instead of incidence breaks that symmetry: with $y=I$, \jlcmd{assess_identifiability} classifies $\beta_I$, $\gamma$, and $\mu$ as globally identifiable, with $\beta_W$ and $\xi$ still confounded through the product $\beta_W\xi$, which a concurrent measurement of $W$ resolves through $\dot W=\xi I-\mu W$. These conclusions are consistent with the analysis of Eisenberg \emph{et al.}~\cite{Eisenberg2013}, who work with a rescaled reservoir variable. Model~\eqref{eq:siwr} is also the cholera model analyzed by Pant \emph{et al.}~\cite{Pant2026}, who write $\alpha$ and $\zeta$ for our $\xi$ and $\mu$. With the pathogen concentration $W(t)$ observed and the initial population size known, they likewise find that $\gamma+\mu$ and $\gamma\mu$ are globally identifiable while $\gamma$ and $\mu$ are only locally identifiable, and they show that single-time measurements of $S$, $I$, or $R$ restore global identifiability (Section~\ref{subsec:reparam-not-enough}).

\paragraph{Discussion.}
This example illustrates how structural identifiability can guide surveillance design in a quantitative way. Incidence data alone determine the direct transmission rate and a mixed combination of the environmental parameters; environmental sampling fixes the shedding rate; and distinguishing recovery from environmental decay requires prevalence information or prior knowledge of one of the two rates, or of which of them is larger, since the two branches of the ambiguity exchange $\gamma$ and $\mu$. Each data stream removes a specific, nameable ambiguity, and the analysis tells the modeler which stream is worth collecting.

\subsection{An SEIR model with hospitalization}\label{subsec:seirh}

\paragraph{Background.}
SEIR models with hospitalization became widely used during the COVID-19 pandemic for forecasting hospital demand and disease transmission. The exposed compartment captures the incubation period, while hospitalization provides information for healthcare planning. Reported hospital admissions and case incidence are the primary data sources used for model calibration~\cite{Liyanage2025}.

\paragraph{Model.}
The SEIR-H model is
\begin{equation}\label{eq:seirh}
\begin{aligned}
\dot S &= -\beta SI,\\
\dot E &= \beta SI-\sigma E,\\
\dot I &= \sigma E-(\gamma+\eta)I,\\
\dot H &= \eta I-\rho H,\\
\dot R &= \gamma I+\rho H,
\end{aligned}
\end{equation}
where $\sigma$ is the progression rate from exposed to infectious, $\eta$ is the hospitalization rate, $\rho$ is the discharge rate, and $\gamma$ is the recovery rate of non-hospitalized individuals. Hospital admissions,
\[
y_1(t)=\eta I(t),
\]
are the primary observed output, with case incidence,
\[
y_2(t)=\beta S(t)I(t),
\]
optionally included as a second measurement.

\paragraph{Julia implementation.}\, \\
\begin{lstlisting}
ode = @ODEmodel(
    S'(t) = -beta*S(t)*I(t),
    E'(t) =  beta*S(t)*I(t) - sigma*E(t),
    I'(t) =  sigma*E(t) - (gamma + eta)*I(t),
    H'(t) =  eta*I(t) - rho*H(t),
    R'(t) =  gamma*I(t) + rho*H(t),
    y1(t) =  eta*I(t)
)
\end{lstlisting}

\paragraph{Results.}
With hospital admissions as the only observed output, \jlcmd{assess_identifiability} classifies $\sigma$ as locally identifiable and every other parameter ($\beta$, $\gamma$, $\eta$, $\rho$) and every state as non-identifiable. The result is less favorable than intuition might suggest, and the reasons are instructive. Because $H$ and $R$ do not feed back into the infection dynamics, nothing about $\rho$ reaches the output; the discharge rate is invisible unless hospital occupancy is measured. In addition, the transformation
\[
(S,E,I)\mapsto\lambda\,(S,E,I),\qquad \beta\mapsto\beta/\lambda,\qquad \eta\mapsto\eta/\lambda,\qquad \gamma\mapsto\gamma+\eta(1-1/\lambda),
\]
leaves $\eta I$ unchanged for every $\lambda>0$: admissions data fix neither the scale of the epidemic nor the split of the exit rate $\gamma+\eta$ into recovery and hospitalization. Eliminating the unobserved states from the equations for $J=\eta I$ yields an input--output equation whose coefficients are $\sigma+\gamma+\eta$, $\sigma(\gamma+\eta)$, and $\beta/\eta$, and \jlcmd{find_identifiable_functions} returns exactly these three quantities as generators of the identifiable functions of this scheme. The first two determine $\sigma$ and $\gamma+\eta$ only up to exchange, which is the source of the local verdict for $\sigma$.

\paragraph{Additional measurements.}
Adding case incidence, $y_2=\beta SI$, fixes the scale of the epidemic: \jlcmd{assess_identifiability} then classifies $\beta$, $\sigma$, $\gamma$, and $\eta$ as locally identifiable and $S$ as observable, while $\rho$ (and $H$, $R$) remain non-identifiable. The remaining ambiguity is discrete. Eliminating $E$ from the equations for $y_1$ and $y_2$ gives
\[
\ddot y_1+(\sigma+\gamma+\eta)\,\dot y_1+\sigma(\gamma+\eta)\,y_1=\sigma\eta\,y_2,
\]
which is invariant under the exchange $\sigma\leftrightarrow\gamma+\eta$ (with $E$ adjusted). The identifiable functions returned for this scheme are $\{\sigma+\gamma+\eta,\ \gamma\sigma,\ \eta\sigma,\ \beta\sigma\}$. This is the classical two-compartment chain ambiguity of Section~\ref{subsec:pk} reappearing inside a nonlinear epidemic model. Two further pieces of information complete the picture. Fixing $\sigma$ from clinical data on the incubation period, a common practice, resolves the exchange and makes $\beta$, $\gamma$, and $\eta$ globally identifiable, since $\eta=\sigma\eta/\sigma$, $\gamma=(\gamma+\eta)-\eta$, and $\beta=(\beta/\eta)\eta$; \jlcmd{assess_identifiability} confirms this after $\sigma$ is fixed with \jlcmd{set_parameter_values}. Observing hospital occupancy $H(t)$, which is routinely reported, identifies $\rho$ directly from $\dot H=y_1-\rho H$, and the software classifies $\rho$ as globally identifiable once $y_3=H$ is added. Extended SEIR models with disease-induced mortality behave similarly, with daily death counts serving as the additional informative measurement~\cite{Liyanage2025}.

\paragraph{Discussion.}
This example demonstrates that structural identifiability depends critically on the available surveillance streams and that the answer can be considerably less favorable than a compartment diagram suggests. Admissions alone identify three parameter combinations; adding incidence resolves the scale but leaves a two-fold ambiguity between incubation and exit rates; fixing the incubation rate or adding occupancy data completes the identification. The analysis thus provides a principled basis for deciding which data streams must be combined, and which prior information must be supplied, before a forecasting model of this type is calibrated.

\subsection{Consolidated verdicts}\label{subsec:consolidated}

\begin{figure}[!ht]
\centering
\includegraphics[width=\linewidth]{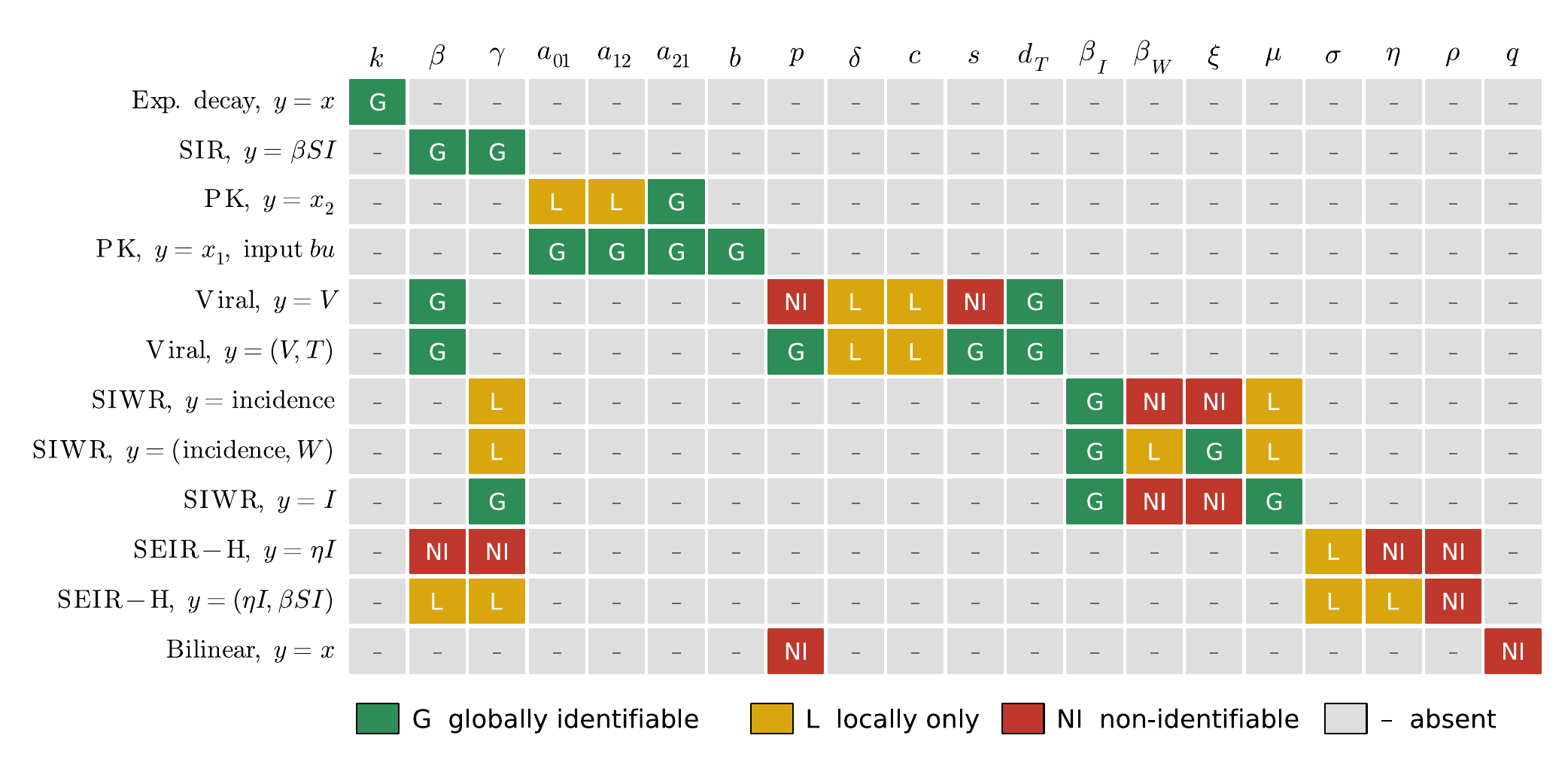}
\caption{\textbf{Structural identifiability across the case studies considered in this tutorial.} Each row corresponds to a model together with its observation scheme; each column is a model parameter (parameters with the same name in different models, such as $\beta$ and $\gamma$, share a column). Green cells denote globally identifiable parameters (\Glob), amber cells parameters that are locally but not globally identifiable (\Loc), red cells non-identifiable parameters (\Nid), and gray cells parameters absent from the model. Reading down a column shows how the same parameter changes status with the measurement scheme; reading along a row shows which ambiguities remain. The matrix conveys the central message of this tutorial: structural identifiability depends jointly on the mathematical model and the measurement scheme, and additional measurements remove specific ambiguities rather than all of them at once.}
\label{fig:verdicts}
\end{figure}

Table~\ref{tab:cases} and Figure~\ref{fig:verdicts} summarize the structural identifiability results across all case studies. Three general observations emerge. First, structural identifiability depends not only on the model but also on the measurement scheme: the pharmacokinetic model ranges from globally identifiable to non-identifiable, and the viral dynamics, SIWR, and SEIR-H models each gain identifiable parameters when a second observable is added. Second, structural non-identifiability is rarely complete. Even when individual parameters cannot be uniquely recovered, identifiable combinations remain, and these need not be simple products or sums: the SIWR generator $\beta_I\mu+\beta_W\xi$ mixes the two transmission pathways, and the viral dynamics generator $ps$ couples a source term with a production rate through a common scaling of two unobserved states. Third, discrete and continuous ambiguities require different remedies. A continuous ambiguity (a scaling symmetry) is removed by a measurement or a known initial condition that fixes the missing scale; a discrete ambiguity (an exchange symmetry) is removed only by information that breaks the symmetry, such as a fixed parameter or an observable that distinguishes the exchanged compartments.

\begin{table}[!ht]
\centering
\caption{Summary of the structural identifiability results obtained for all case studies considered in this tutorial. For each model and observation scheme, the table reports the number of state variables, free parameters, observed outputs, the overall verdict, a generating set of identifiable functions or the parameter-level verdicts, and the corresponding section. All initial conditions are unknown. Verdict abbreviations: \Glob{} = globally identifiable; \Loc{} = locally but not globally identifiable; \Nid{} = non-identifiable. ``Partial'' indicates that different parameters fall into different identifiability classes. Generating sets marked with $^{\ast}$ were returned by \jlcmd{find_identifiable_functions}; the remaining entries list the parameter-level verdicts of \jlcmd{assess_identifiability}.}
\label{tab:cases}
\renewcommand{\arraystretch}{1.25}
\setlength{\tabcolsep}{4pt}
\footnotesize
\begin{tabular}{@{}>{\raggedright\arraybackslash}p{3.1cm}
c
c
>{\raggedright\arraybackslash}p{2.3cm}
>{\raggedright\arraybackslash}p{1.9cm}
>{\raggedright\arraybackslash}p{4.9cm}
c@{}}
\toprule
\rowcolor{tabhead}
\textbf{Model} &
\textbf{States} &
\textbf{Params} &
\textbf{Observed outputs} &
\textbf{Verdict} &
\textbf{Identifiable functions / parameter verdicts} &
\textbf{\S} \\
\midrule

Exponential decay
& 1 & 1
& $y=x$
& \Glob
& $\{k\}^{\ast}$
& \ref{subsec:expdecay} \\

SIR
& 3 & 2
& $y=\beta SI$
& \Glob
& $\{\beta,\gamma\}$
& \ref{subsec:sir} \\

Two-compartment PK
& 2 & 3
& $y=x_2$, input $u$
& Partial (\Glob/\Loc)
& $\{a_{21},\,a_{01}+a_{12},\,a_{01}a_{12}\}^{\ast}$
& \ref{subsec:pk} \\

Two-compartment PK (central)
& 2 & 4
& $y=x_1$, input $bu$
& \Glob
& $\{a_{01},a_{12},a_{21},b\}^{\ast}$
& \ref{subsec:pk} \\

Viral dynamics
& 3 & 6
& $y=V$
& Partial
& $\{d_T,\,\beta,\,c+\delta,\,c\delta,\,ps\}^{\ast}$
& \ref{subsec:viral} \\

Viral dynamics (augmented)
& 3 & 6
& $(V,T)$
& Partial (\Glob/\Loc)
& $s,d_T,\beta,p$: \Glob; $c,\delta$: \Loc
& \ref{subsec:viral} \\

SIWR
& 4 & 5
& Incidence
& Partial
& $\{\beta_I,\,\gamma+\mu,\,\gamma\mu,\,\beta_I\mu+\beta_W\xi\}^{\ast}$
& \ref{subsec:siwr} \\

SIWR (augmented)
& 4 & 5
& Incidence, $W$
& Partial (\Glob/\Loc)
& $\beta_I,\xi$: \Glob; $\beta_W,\gamma,\mu$: \Loc
& \ref{subsec:siwr} \\

SIWR (prevalence)
& 4 & 5
& $y=I$
& Partial
& $\{\beta_I,\,\gamma,\,\mu,\,\beta_W\xi\}^{\ast}$
& \ref{subsec:siwr} \\

SEIR--H
& 5 & 5
& $y=\eta I$
& Partial (\Loc/\Nid)
& $\{\sigma+\gamma+\eta,\,\sigma(\gamma+\eta),\,\beta/\eta\}^{\ast}$
& \ref{subsec:seirh} \\

SEIR--H (augmented)
& 5 & 5
& $(\eta I,\beta SI)$
& Partial (\Loc/\Nid)
& $\beta,\sigma,\gamma,\eta$: \Loc; $\rho$: \Nid
& \ref{subsec:seirh} \\

Bilinear toy model
& 1 & 2
& $y=x$
& \Nid
& $\{pq\}^{\ast}$
& \ref{sec:reparam} \\

\bottomrule
\end{tabular}
\end{table}

\section{Reparameterization and Identifiable Combinations}\label{sec:reparam}

\subsection{The canonical bilinear example}

\begin{figure}[!ht]
  \centering
  \includegraphics[width=\linewidth]{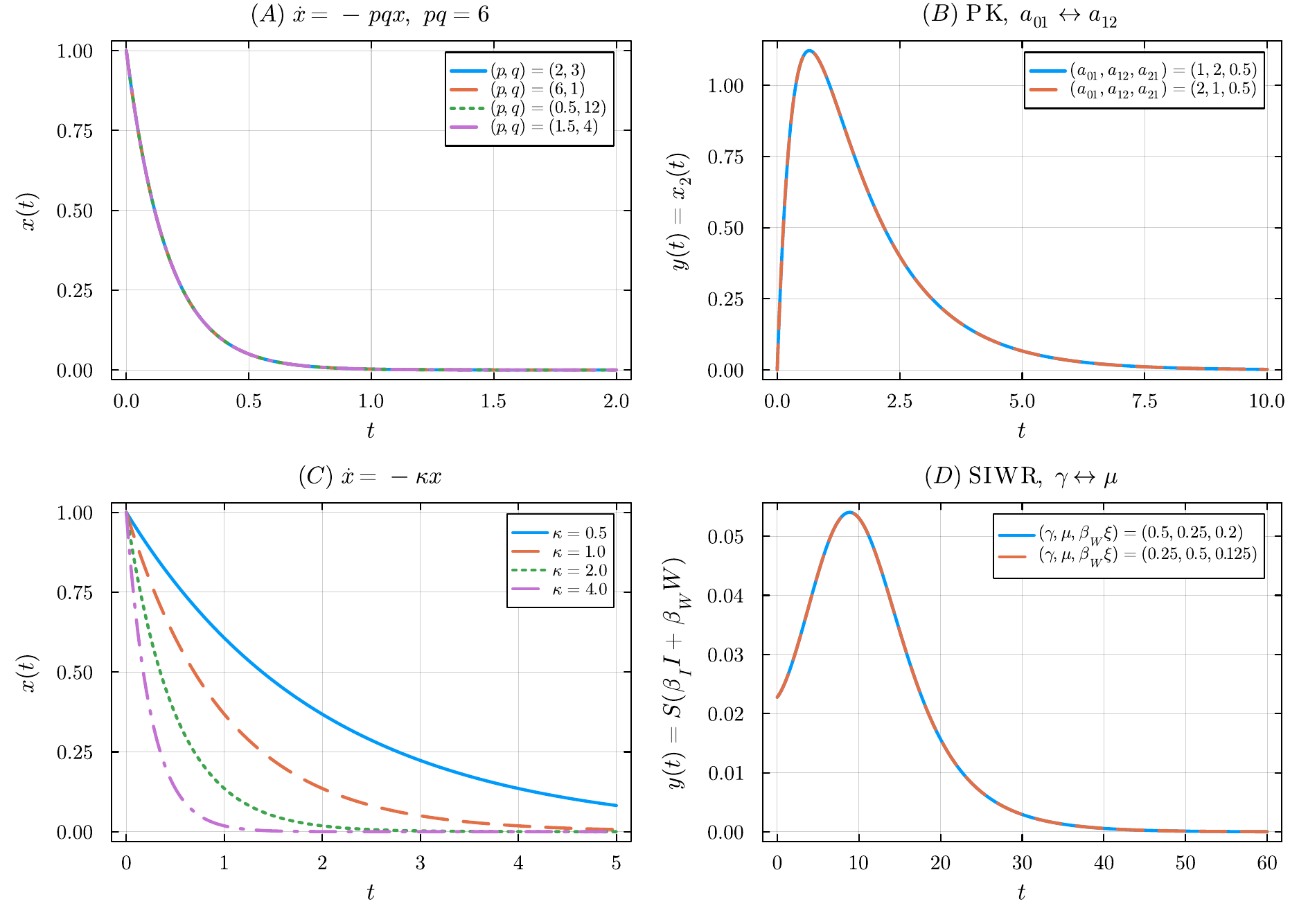}
  \caption{\textbf{Numerical demonstration of structural non-identifiability and its remedy.} (A) Bilinear toy model $\dot x = -p q\, x$ integrated for four distinct pairs $(p, q) = (2,3), (6,1), (0.5,12), (1.5,4)$, all sharing the product $pq = 6$. The four trajectories coincide, illustrating that only the product is identifiable. (B) Two-compartment pharmacokinetic model of Section~\ref{subsec:pk} with the peripheral compartment observed, integrated after a bolus dose $x_1(0)=10$, $x_2(0)=0$, for the two parameter sets related by the $a_{01}\leftrightarrow a_{12}$ exchange, $(a_{01},a_{12},a_{21}) = (1,2,0.5)$ and $(2,1,0.5)$. The two output trajectories $x_2(t)$ agree to within roundoff error (maximum absolute difference $1.1\times10^{-13}$ with relative and absolute solver tolerances of $10^{-12}$), confirming that $a_{01}$ and $a_{12}$ are locally but not globally identifiable. (C) After the reparameterization $\kappa:=pq$, distinct values of $\kappa$ produce distinct trajectories: the reformulated model is globally identifiable. (D) SIWR model of Section~\ref{subsec:siwr} with incidence observed, for $\beta_I=0.3$ and the parameter sets $(\gamma,\mu,\beta_W\xi)=(0.5,0.25,0.2)$ and $(0.25,0.5,0.125)$ related by the transformation of Section~\ref{subsec:siwr}, with initial states $(S,I,\beta_WW)=(0.99,0.01,0.02)$ and $(0.99,0.035,0.0125)$. The incidence curves agree to within $7.5\times10^{-14}$, although the two parameter sets differ in all three of $\gamma$, $\mu$, and $\beta_W\xi$.}
  \label{fig:nonidentifiability}
\end{figure}

We pause the sequence of realistic models to examine, with maximal clarity, the simplest nontrivial structural non-identifiability. Consider the bilinear scalar system
\begin{equation}\label{eq:bilinear}
\dot x(t) = -p\,q\,x(t),\qquad y(t) = x(t),
\end{equation}
with $\theta = (p, q)$ and $x(0)$ known. The closed form solution $x(t) = x(0)\, e^{-pqt}$ depends on $p$ and $q$ only through their product. The pair $(p, q) = (2, 3)$ and the pair $(p, q) = (6, 1)$ generate identical output trajectories; no experiment can distinguish them. Figure~\ref{fig:nonidentifiability}(A) provides a direct numerical verification: four distinct $(p, q)$ pairs sharing $pq = 6$ collapse onto a single output curve.

\jlcmd{StructuralIdentifiability.jl} confirms the algebraic statement immediately:
\begin{lstlisting}
ode = @ODEmodel(
    x'(t) = -p * q * x(t),
    y(t)  =  x(t)
)
assess_identifiability(ode)
# p => :nonidentifiable
# q => :nonidentifiable
find_identifiable_functions(ode)
# returns [p*q]
\end{lstlisting}

The model is non-identifiable in $p$ and $q$, yet the product $pq$ is globally identifiable. The reparameterization is therefore canonical: introducing $\kappa := pq$ yields
\begin{equation*}
\dot x(t) = -\kappa\, x(t),
\end{equation*}
which is precisely the exponential decay model of Section~\ref{subsec:expdecay}. The reformulated model contains one parameter, is globally identifiable, and captures exactly the information that the data can deliver.

The analogous phenomenon for the two-compartment pharmacokinetic model is illustrated in Figure~\ref{fig:nonidentifiability}(B). Two parameter triples that exchange the values of $a_{01}$ and $a_{12}$ while preserving their sum and product produce peripheral-compartment trajectories that are indistinguishable to numerical precision, from the same initial condition. The verdict of Section~\ref{subsec:pk}, namely that $a_{01}$ and $a_{12}$ are locally but not globally identifiable, is thus borne out by direct simulation. Such simulations are a useful sanity check on a probabilistic symbolic verdict and take seconds to run.

\subsection{Identifiable combinations as a constructive object}

The bilinear example crystallizes a general principle. Whenever \jlcmd{find_identifiable_functions} returns generators that are nontrivial functions of the original parameters, those generators constitute the model's appropriate parameter vector. Three immediate consequences follow.
\begin{enumerate}[leftmargin=*,itemsep=2pt]
  \item \textbf{Fitting.} Likelihood, Bayesian, and least squares estimation should be performed in the new coordinates $\eta = \phi(\theta)$ rather than in $\theta$. Optimization on $\theta$ is ill posed, whereas optimization on $\eta$ is well posed by construction.
  \item \textbf{Reporting.} Confidence intervals and posterior credible regions should be reported on $\eta$. Intervals on $\theta$ in the absence of identifiability are not interpretable; they reflect the prior or the regularizer~\cite{Raue2009}.
  \item \textbf{Comparing studies.} Two studies that independently estimate $\theta$ from non-identifiable models cannot be combined meaningfully, whereas two studies that estimate $\eta$ can.
\end{enumerate}

\subsection{When reparameterization is not enough}\label{subsec:reparam-not-enough}

There exist settings in which the identifiable functions are scientifically uninterpretable, for instance when a sum of dissimilar rate constants whose individual values are biologically meaningful turns out to be the only identifiable combination. In such cases the available remedies are:
\begin{itemize}[leftmargin=*,itemsep=2pt]
  \item \textbf{Add measurements.} If the identifiable field is too small, additional outputs (a second compartment or an environmental measurement) enlarge it, exactly as in Sections~\ref{subsec:viral}, \ref{subsec:siwr}, and \ref{subsec:seirh} above.
  \item \textbf{Combine experimental conditions.} Experiments under different conditions that share some parameters can enlarge the identifiable field, as in the two-condition example of Section~\ref{subsec:limitations}. Repeating an experiment of the same design with new initial conditions helps only when the single-experiment ambiguity depends on the initial state, as in Example~2.14 of Section~\ref{subsec:io}; it leaves the pharmacokinetic exchange symmetry unresolved (Section~\ref{subsec:limitations}).
  \item \textbf{Fix parameters or initial conditions from external data.} If a parameter is known from an independent experiment, it can be removed from the parameter vector, and the remaining parameters may become identifiable. A known initial condition acts in the same way (Section~\ref{subsec:viral}).
  \item \textbf{Add inputs.} A controllable input $u(t)$ can substantially enlarge the identifiable field. This is the principle behind input design in pharmacokinetics~\cite{Walter1997}.
  \item \textbf{Use qualitative prior information to select a branch.} When the ambiguity is discrete, information from outside the data, such as evidence that one process is faster than another, can select one of the finitely many parameter sets that produce the same output (Sections~\ref{subsec:viral} and~\ref{subsec:siwr}). Inequality constraints of this kind cannot remove a continuous ambiguity; they only bound it. Prior distributions in Bayesian estimation likewise leave structural identifiability unchanged: along a non-identifiable direction, the posterior reflects the prior~\cite{Raue2009}.
\end{itemize}
The unifying theme is that identifiability is a property of both the model and the experiment, and the modeler has leverage on both. These remedies change the structural problem; whether the resulting parameters can be estimated precisely from finite, noisy data is the separate question of practical identifiability (Section~\ref{subsec:practical-vs-structural}).

A complete second output trajectory is the most demanding of these remedies, and it is not always needed. A measurement of a complementary state at a single time point adds one algebraic equation, $x_j(t_1)=c$, that relates the unknown parameters and initial conditions to a known number. When the ambiguity is a one-dimensional continuous family, such as the scaling symmetries of Sections~\ref{subsec:viral} and~\ref{subsec:seirh}, one such equation can be enough to single out one member of the family. Pant \emph{et al.}~\cite{Pant2026} prove results of this kind for several epidemic models. For the SIR model with frequency-dependent transmission $\beta SI/N$, observed through incidence and with an unknown population size $N$, the value of $R$ at one time point makes the model globally identifiable (their Theorem~3.1). For a cholera model of the form~\eqref{eq:siwr} with a known initial population size, observed through the pathogen concentration $W(t)$, one value of each of two different compartments among $S$, $I$, and $R$ is required, whereas observing the shedding inflow $\xi I(t)$ instead reduces the requirement to a single value of any one of them (their Theorems~3.4 and~3.5). The number and type of point measurements needed are therefore specific to the model and to the continuously observed output, and they must be established for the model at hand. A known initial condition, as in Section~\ref{subsec:viral}, is the special case $t_1=0$. These results assume noise-free measurements, so their practical value depends on how precisely the single value can be measured.

Sampling an already measured output more densely is a different matter. Structural identifiability assumes that this output is known as a function of time, so additional samples of it can improve practical identifiability but cannot change a structural verdict, whether the ambiguity is continuous or discrete. More generally, the input $u(t)$, the sampling schedule, and the experimental conditions can be chosen before data collection to make the data as informative as possible about a target set of parameters or identifiable combinations~\cite{Walter1997}. Structural identifiability analysis determines \emph{whether} an experiment can succeed in principle; optimal experimental design determines \emph{which} of the structurally admissible experiments is statistically most efficient. The appropriate remedy depends on the cost, in a given application, of instrumenting a new data stream relative to taking a single additional measurement or redesigning the sampling of an existing one.

\section{Best Practices and Recommendations}\label{sec:best}

This section summarizes practical recommendations that emerge from the case studies.

\subsection{Assess identifiability before parameter estimation}

Structural identifiability analysis should precede parameter estimation. Fitting a structurally non-identifiable model may produce an apparently optimal solution together with plausible confidence intervals, even though infinitely many parameter sets generate identical model outputs. In such cases, the reported parameter estimates are artifacts of the optimization procedure or regularization rather than properties of the underlying model~\cite{Raue2009,Wieland2021}.

For the six case-study models timed by the companion script \texttt{12\_timing.jl} (Table~\ref{tab:scripts}), \jlcmd{assess_local_identifiability} completed in less than 0.02~s and \jlcmd{assess_identifiability} in less than 0.2~s on a laptop (Intel Core i9-12900H, Julia~1.12.2, timed after compilation). Such times are typical of small models but should not be extrapolated, because the cost of both tests grows with the size and the algebraic complexity of the model (Section~\ref{subsec:limitations}). The local test remains the natural first diagnostic, since it is usually much cheaper than the global analysis.

\subsection{Perform local analysis before global analysis}

A practical workflow is to evaluate local identifiability before performing global analysis. The local test rapidly detects structurally non-identifiable parameters, eliminating the need for a more computationally demanding global analysis of those parameters; identifiable combinations that involve them are then obtained with \jlcmd{find_identifiable_functions}. When all parameters are locally identifiable, \jlcmd{assess_identifiability} can then determine whether they are globally identifiable or only locally identifiable because of discrete parameter symmetries.

\subsection{Treat the measurement scheme as part of the model}

Structural identifiability depends jointly on the model equations and the measurement scheme. A model that is non-identifiable under one set of observations may become globally identifiable after additional outputs are measured. The case studies illustrate this principle in both directions: observing the central instead of the peripheral compartment turns the pharmacokinetic model from locally into globally identifiable; measuring the target-cell population $T(t)$ removes the scaling ambiguity of the viral dynamics model; and measuring the environmental pathogen concentration $W(t)$ identifies the shedding rate in the SIWR model, whereas prevalence rather than incidence is needed to separate recovery from environmental decay. Consequently, selecting informative measurements should be considered an integral part of model design rather than an experimental afterthought.

\subsection{Specify initial conditions carefully}

Assumptions regarding the initial conditions can substantially affect structural identifiability. Treating initial conditions as known, unknown, or additional parameters may change the identifiability status of several model parameters. Accordingly, identifiability analyses should explicitly state these assumptions and, whenever appropriate, compare the results obtained under known and unknown initial conditions~\cite{Liyanage2025}; in the viral dynamics model, declaring $T(0)$ known removes the scaling ambiguity between $s$ and $p$ (Section~\ref{subsec:viral}). Section~\ref{subsec:phi} distinguishes four such assumptions. In particular, the \jlcmd{known_ic} option treats the declared values as generic, whereas specific values such as a zero initial drug concentration define a different problem that must be examined separately.
\subsection{Report identifiable combinations}

When \jlcmd{find_identifiable_functions} identifies parameter combinations, these should be reported instead of the original parameters. For example, fitting the viral dynamics model of Section~\ref{subsec:viral} to viral load data alone cannot estimate $s$ and $p$ individually, nor separate $c$ from $\delta$, but it can identify $ps$, $c+\delta$, and $c\delta$~\cite{Miao2011}. Reporting identifiable combinations provides statistically and biologically meaningful results, whereas interpreting non-identifiable parameters individually can be misleading.

\subsection{Distinguish structural and practical identifiability}\label{subsec:practical-vs-structural}

Structural identifiability is a prerequisite for parameter estimation but does not guarantee practical identifiability. Even when a parameter is structurally identifiable, it may be estimated unreliably because of limited or noisy data, poor experimental design, or weak parameter sensitivity~\cite{Raue2009,Wieland2021}. After establishing structural identifiability, practical identifiability should be evaluated using complementary methods such as profile likelihoods, the Fisher Information Matrix, or Markov chain Monte Carlo diagnostics. In the Julia ecosystem, \jlcmd{LikelihoodProfiler.jl}~\cite{Borisov2020LikelihoodProfiler} provides profile-likelihood analysis for assessing practical identifiability.

\subsection{Common mistakes to avoid}

The following practices frequently lead to incorrect parameter estimation or misleading interpretation:

\begin{itemize}[leftmargin=*,itemsep=2pt]
\item Performing parameter estimation before assessing structural identifiability.

\item Assuming that optimization algorithms uniquely estimate structurally non-identifiable parameters, including parameters related by exchange symmetries or other model symmetries~\cite{Cobelli1980}.

\item Specifying initial conditions without explicitly stating whether they are known, estimated, or assumed.

\item Interpreting structural non-identifiability as a numerical or software issue rather than an intrinsic property of the model and measurement scheme.

\item Failing to repeat the identifiability analysis after modifying the model equations, parameters, or observed outputs.

\item Transcribing an identifiability verdict from a textbook or documentation example to a model with a different measurement scheme. As Section~\ref{subsec:pk} shows, the same two-compartment model is locally identifiable when the peripheral compartment is observed and globally identifiable when the central compartment is observed.
\end{itemize}

\subsection{Limitations of structural identifiability}\label{subsec:limitations}

Structural identifiability analysis is an essential diagnostic tool, but it has important limitations. This section discusses the computational factors affecting scalability, the current treatment of multiple experimental conditions, the scope of the rational-model assumption, and issues beyond structural identifiability, including measurement noise and model misspecification. It concludes with practical recommendations for applying structural identifiability analysis effectively.

\paragraph{Computational complexity and scalability.}

The cost of an identifiability analysis is not determined by the number of states alone. It depends on the numbers of states and unknown parameters, on the number and form of the outputs and inputs, on the degrees of the polynomials in the model, on the order and the number of terms of the resulting input--output equations, and on the question being asked: the local test is usually the cheapest, the global test is more expensive, and the computation and simplification of identifiable functions can be more expensive still. The cost of Sedoglavic's local test is bounded by a polynomial in these quantities~\cite{Sedoglavic2002}, but a polynomial bound does not make the cost negligible, since it grows with the numbers of states, parameters, and outputs and with the degree of the equations. The global analysis has no comparable general bound: differential elimination can produce input--output equations with very many terms, and the runtime is hard to predict from the size of the model.

The benchmarks of Dong \emph{et al.}~\cite{Dong2023} (their Tables~2 and~3) illustrate this. With their implementation, the global analysis of a pharmacokinetic model with four states and seven parameters took about 400~s, that of a MAPK pathway model with twelve states and 22 parameters about 40~s, and that of an NF-$\kappa$B model with fifteen states did not finish within five hours. These timings depend on the hardware and on the software version, and the package has been developed further since then, but the absence of a monotone relation between model size and runtime is the general point. For the timed models of this tutorial, each with at most five states, the local and global tests completed in a fraction of a second (Section~\ref{sec:best}). The benchmark file of the package repository (\texttt{benchmarking/benchmarks.jl}) currently lists 65 published models, including MAPK, JAK--STAT, p53, and NF-$\kappa$B signaling models, on which the global algorithms are tested. We therefore do not state a model size beyond which global analysis becomes infeasible. In practice, we recommend the following order: run \jlcmd{assess_local_identifiability} first; then run \jlcmd{assess_identifiability} on the actual model with a time limit; and, if the full global analysis is too expensive, analyze the submodels returned by \jlcmd{find_submodels}, which the package suggests when a model has nontrivial submodels. Restricting the question with \jlcmd{funcs_to_check} does not restrict the elimination: in version~0.5.26, the input--output equations of the whole model are computed in every case, and only the subsequent membership test depends on the requested functions. By default, \jlcmd{assess_identifiability} also reports the states and therefore computes, in addition, the generators that involve states; checking only functions of the parameters skips that step. When the requested functions involve states, \jlcmd{assess_identifiability} with or without \jlcmd{funcs_to_check} and \jlcmd{find_identifiable_functions} with \jlcmd{with_states=true} carry out essentially the same elimination and have comparable costs, the latter adding a final simplification of the generators. Users are encouraged to report the computational environment together with runtime measurements.

\paragraph{Multiple experimental conditions.}

Several situations that are informally described as ``more data'' lead to different identifiability problems.
\begin{enumerate}[label=(\alph*),leftmargin=*,itemsep=2pt]
\item \emph{Several outputs measured in one experiment} add output equations to a single model, as in Sections~\ref{subsec:viral}--\ref{subsec:seirh}. They are supported directly.
\item \emph{Several known inputs in one experiment} are declared as \jlcmd{u1(t)}, \jlcmd{u2(t)}, \ldots{} and are treated as known generic functions of time. They are supported directly.
\item \emph{Repeated experiments with the same design}, that is, with the same parameters but independent initial conditions, define multi-experiment identifiability. \jlcmd{StructuralIdentifiability.jl} reports it for the local question through \jlcmd{type = :ME} (Section~\ref{subsec:api}), and its global verdicts refer to it whenever the Wronskian check fails (Section~\ref{subsec:io}).
\item \emph{Experiments under different conditions}, such as different doses, interventions, or environments, involve \emph{shared parameters}, which take the same value in every condition, and \emph{condition-specific parameters} or inputs, which differ between conditions.
\end{enumerate}
The package has no dedicated interface for case~(d). Such designs can be analyzed by building one augmented model that contains a copy of the states for each condition, each copy with its own unknown initial conditions, in which shared parameters appear once and condition-specific parameters and inputs receive separate names. For the two-compartment model of Section~\ref{subsec:pk} with the peripheral compartment observed, two experiments with the same design leave $a_{01}$ and $a_{12}$ locally identifiable, because both experiments share the same exchange symmetry. Suppose instead that the second experiment changes the elimination rate, for example through an inhibitor of elimination, so that its elimination rate $b_{01}$ is condition-specific while $a_{12}$ and $a_{21}$ are shared:
\begin{lstlisting}
ode = @ODEmodel(
    x1'(t) = -(a01 + a21) * x1(t) + a12 * x2(t) + u1(t),  # condition A
    x2'(t) =  a21 * x1(t) - a12 * x2(t),
    z1'(t) = -(b01 + a21) * z1(t) + a12 * z2(t) + u2(t),  # condition B
    z2'(t) =  a21 * z1(t) - a12 * z2(t),
    yA(t)  =  x2(t),
    yB(t)  =  z2(t)
)
assess_identifiability(ode; funcs_to_check = [a01, b01, a12, a21])
# a01 => :globally, b01 => :globally, a12 => :globally, a21 => :globally
\end{lstlisting}
All four rates become globally identifiable. Each condition determines the sum and the product of its own elimination rate and $a_{12}$, so $a_{12}$ is the root that the two resulting quadratic equations have in common. The augmented model has as many states as all conditions together, so the cost of the analysis grows with the number of conditions. For parameter estimation, \jlcmd{PEtab.jl}~\cite{Persson2024PEtab} handles multiple simulation conditions natively (Section~\ref{sec:ecosystem}), and among identifiability tools GenSSI~2.0 was designed for multi-experiment analysis of SBML models~\cite{Ligon2018}.

\paragraph{Non-rational nonlinearities.}

The algorithms of \jlcmd{StructuralIdentifiability.jl} require rational right-hand sides, and the package stops with an error message when it detects a non-rational term. Non-rational models are not thereby excluded from symbolic analysis: exponential, logarithmic, and power-law terms can often be removed by introducing new state variables, after which the standard workflow applies. The way in which the new variables are introduced matters, however. Liyanage \emph{et al.}~\cite{LiyanageViruses2025} analyze the generalized growth model $\dot C=rC^{\alpha}$ with incidence $y=\dot C$ observed. Setting $x=rC^{\alpha}$ gives the rational system $\dot C=x$, $\dot x=\alpha x^{2}/C$, $y=x$, for which \jlcmd{StructuralIdentifiability.jl} classifies $\alpha$, $C$, and $x$ as globally identifiable, so that $r=x/C^{\alpha}$ is identifiable as well, in agreement with their Proposition~1. Setting instead $w=C^{\alpha}$ gives $\dot C=rw$, $\dot w=\alpha rw^{2}/C$, $y=rw$, and the package reports $r$ as non-identifiable, with only the product $rw$ identifiable. The second lifting has more trajectories than the original model, because the relation $w(0)=C(0)^{\alpha}$ is not imposed, and the extra trajectories create the spurious symmetry $(r,w)\mapsto(\lambda r,w/\lambda)$~\cite{LiyanageViruses2025}. In the first lifting, every trajectory of the lifted model corresponds to a trajectory of the original one: the new initial value $x(0)=rC(0)^{\alpha}$ simply takes the place of $r$ among the unknowns. The supplementary material of the SIAN paper~\cite{Hong2019SIAN} (Sections~A.2 and~A.3) gives two further examples. In a model of ruminal lipolysis, a time-dependent factor $k_1e^{-k_3t}$ is replaced by a state $x_5$ with $\dot x_5=-k_3x_5$ whose initial value takes the place of $k_1$, and in the Goodwin oscillator a Hill term with a non-integer exponent is treated in the same way; the documentation of \jlcmd{StructuralIdentifiability.jl} analyzes the Goodwin oscillator in this lifted form. Models augmented with neural networks can often be lifted in the same way. For a hidden unit with activation $\tanh$, the variable $h=\tanh(w^{\top}x+b)$ satisfies $\dot h=(1-h^{2})\,w^{\top}\dot x$, which is rational in $h$ and $x$, and its initial value takes the place of the bias $b$ among the unknowns; the logistic sigmoid $\sigma$, with $\sigma'=\sigma(1-\sigma)$, and the exponential behave in the same way, and deeper networks are lifted layer by layer. The lifted model has one new state per hidden unit, so exact lifting of a large network can make the analysis expensive. Activations without such a differential equation, for example the piecewise-linear ReLU, and other piecewise-defined models do not admit a lifting of this kind, and for them approximate or numerical identifiability methods are needed.

\paragraph{Noise and model misspecification.}

Structural identifiability assumes noise-free observations and a correct mathematical model. Consequently, it does not account for measurement noise, limited data, or model misspecification. A parameter may be structurally identifiable yet practically difficult to estimate from noisy observations, while a structurally identifiable model may still fail to describe the underlying biological process adequately. Structural identifiability should therefore be complemented by practical identifiability analysis and independent model validation~\cite{Raue2009}.

\paragraph{Practical recommendations.}

The case studies presented throughout this tutorial suggest the following practical guidelines.

\begin{itemize}[leftmargin=*,itemsep=2pt]
\item Assess local identifiability before performing global analysis or parameter estimation.

\item Simplify the model whenever possible by removing redundant states or parameters.

\item Use identifiable parameter combinations rather than non-identifiable individual parameters for estimation and reporting (Section~\ref{sec:reparam}).

\item Incorporate reliable prior biological knowledge by fixing well-established parameters whenever appropriate.

\item Design measurement schemes and experiments to maximize structural identifiability before collecting data.
\end{itemize}

These considerations should guide the application of structural identifiability analysis rather than discourage its use.

\subsection{Future directions}\label{subsec:future}

Several important challenges remain in symbolic structural identifiability analysis. These include extending current methods to delay differential equations, stochastic differential equations, partial differential equation models, and agent-based models. Another important direction is native support for joint identifiability analysis across multiple experimental conditions with shared model parameters.

Continued integration of \jlcmd{StructuralIdentifiability.jl} with the Julia SciML ecosystem also offers promising opportunities. Closer interoperability with automatic differentiation, parameter estimation, and uncertainty quantification, and identifiability analysis of hybrid mechanistic--machine learning models such as universal differential equations and neural ODEs~\cite{rackauckas2020universal,Chen2018,Norden2025}, will further expand the applicability of structural identifiability analysis in computational biology and biomedical modeling.
\section{Next Steps in the Julia Modeling Workflow}\label{sec:ecosystem}

\begin{figure}[!ht]
\centering
\includegraphics[width=0.95\linewidth]{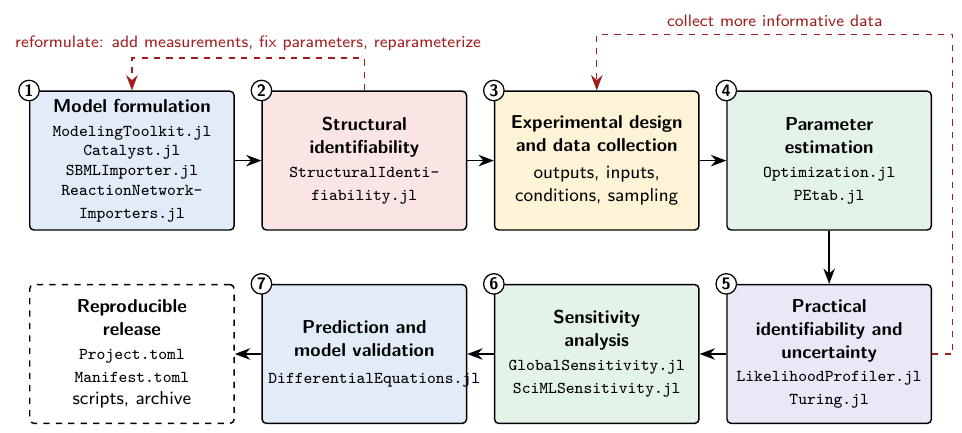}
\caption{\textbf{Structural identifiability within the Julia modeling workflow.} The numbered steps follow Section~\ref{sec:ecosystem}: model formulation, structural identifiability analysis, experimental design and data collection, parameter estimation, practical identifiability and uncertainty quantification, sensitivity analysis, and prediction and model validation, followed by a reproducible release. Dashed arrows show the two most common iterations: a structurally non-identifiable model is reformulated by adding measurements, fixing parameters, or reparameterizing, and a practically non-identifiable model calls for more informative data.}
\label{fig:sciml-workflow}
\end{figure}

Structural identifiability analysis is one step in a longer workflow that begins with model formulation and ends with prediction and model validation. This section describes the steps that usually follow it and the Julia packages that support them. Figure~\ref{fig:sciml-workflow} shows the sequence, and Table~\ref{tab:ecosystem} lists the packages. The sequence is not strictly linear: a structural analysis often sends the modeler back to the model or to the experimental design, and a practical identifiability analysis can do the same.

\paragraph{Model formulation.}
\jlcmd{ModelingToolkit.jl} provides symbolic, equation-based model definitions from which efficient simulation code is generated~\cite{Ma2021MTK}, and \jlcmd{Catalyst.jl} provides a domain-specific language for chemical reaction networks from which deterministic, stochastic, and jump models are derived~\cite{Loman2023Catalyst}. Models stored in standard formats are imported as Catalyst reaction systems: SBML files with \jlcmd{SBMLImporter.jl}~\cite{Bostrom2024SBMLImporter}, and BioNetGen or matrix-based network descriptions with \jlcmd{ReactionNetworkImporters.jl}~\cite{ReactionNetworkImporters} (Section~\ref{subsec:mtk}).

\paragraph{Structural identifiability.}
\jlcmd{StructuralIdentifiability.jl} is applied to the model and measurement scheme following Table~\ref{tab:pipeline}. A non-identifiable result leads back to the model, through reparameterization or fixed parameters (Section~\ref{sec:reparam}), or forward to the experimental design, through additional measurements or conditions.

\paragraph{Experimental design and data collection.}
The identifiability analysis indicates which outputs, inputs, experimental conditions, or single additional measurements (Section~\ref{subsec:reparam-not-enough}) are needed for the quantities of interest to be determined in principle. Choosing among structurally adequate designs, for example sampling times or doses, is a question of optimal experimental design~\cite{Walter1997}.

\paragraph{Parameter estimation.}
\jlcmd{Optimization.jl} provides a common interface to many local and global optimization algorithms~\cite{dixit2023optimization}. \jlcmd{PEtab.jl}~\cite{Persson2024PEtab} builds parameter-estimation problems for Catalyst, ModelingToolkit, and SBML models with multiple observables, multiple simulation conditions, events, and pre-equilibration, reads problems in the PEtab exchange format, and computes gradients by forward-mode automatic differentiation or adjoint sensitivity analysis. The numerical solutions themselves come from the solvers of \jlcmd{DifferentialEquations.jl}~\cite{Rackauckas2017}. Estimation should target identifiable parameters or identifiable combinations (Section~\ref{sec:reparam}).

\paragraph{Practical identifiability and uncertainty quantification.}
Structural identifiability ensures that parameters are determined by ideal data; whether they can be estimated with useful precision from finite, noisy, and discretely sampled data is a separate question~\cite{Raue2009,Wieland2021}. \jlcmd{LikelihoodProfiler.jl} computes profile-likelihood confidence intervals~\cite{Borisov2020LikelihoodProfiler} and is integrated in \jlcmd{PEtab.jl}. \jlcmd{Turing.jl}~\cite{Ge2018Turing} provides Bayesian inference through probabilistic programming, and \jlcmd{DiffEqBayes.jl}~\cite{DiffEqBayes} offers a simplified interface to Bayesian estimation of differential equation models. A structurally non-identifiable parameter appears at this step as a flat likelihood profile or as a posterior distribution stretched along the non-identifiable fiber, but a structurally identifiable parameter can also have a wide confidence interval when the data are insufficient.

\paragraph{Sensitivity analysis.}
\jlcmd{GlobalSensitivity.jl}~\cite{Dixit2022GlobalSensitivity} implements global methods, including the Sobol, Morris, and eFAST methods, that quantify how much of the variation in a model output or prediction is due to each parameter over a range of values, and \jlcmd{SciMLSensitivity.jl}~\cite{rackauckas2020universal} provides forward and adjoint sensitivities of ODE solutions that are also used for gradient-based estimation. Sensitivity analysis helps to decide which identifiable parameters matter for a given prediction and which can be fixed at nominal values.

\paragraph{Prediction and model validation.}
The calibrated model is simulated with \jlcmd{DifferentialEquations.jl} to make predictions, which should be compared with data that were not used for calibration. A prediction that depends on the parameters only through identifiable combinations is determined by the data even when individual parameters are not; Pant \emph{et al.}~\cite{Pant2026} show that this is the case for the basic reproduction number in a range of epidemic models. A prediction that depends on non-identifiable parameters is not. Finally, the scripts, the project environment, and an archived release should be shared (Section~\ref{sec:reproducibility}).

\begin{table}[!ht]
\centering
\caption{Julia packages referenced in this tutorial, organized by step of the modeling workflow.}
\label{tab:ecosystem}
\renewcommand{\arraystretch}{1.10}
\setlength{\tabcolsep}{5pt}
\small
\begin{tabularx}{\linewidth}{@{}l l X@{}}
\toprule
\rowcolor{tabhead}
\textbf{Step} & \textbf{Package} & \textbf{Primary role} \\
\midrule

Model formulation
& \jlcmd{ModelingToolkit.jl}~\cite{Ma2021MTK}
& Symbolic modeling and code generation \\

& \jlcmd{Catalyst.jl}~\cite{Loman2023Catalyst}
& Reaction-network modeling \\

& \jlcmd{SBMLImporter.jl}~\cite{Bostrom2024SBMLImporter}
& SBML model import \\

& \jlcmd{ReactionNetworkImporters.jl}~\cite{ReactionNetworkImporters}
& BioNetGen and matrix import \\

Structural identifiability
& \jlcmd{StructuralIdentifiability.jl}~\cite{Dong2023}
& Symbolic structural identifiability analysis \\

Parameter estimation
& \jlcmd{Optimization.jl}~\cite{dixit2023optimization}
& Optimization framework \\

& \jlcmd{PEtab.jl}~\cite{Persson2024PEtab}
& Multi-condition estimation; PEtab format \\

& \jlcmd{DifferentialEquations.jl}~\cite{Rackauckas2017}
& Numerical simulation \\

Practical identifiability
& \jlcmd{LikelihoodProfiler.jl}~\cite{Borisov2020LikelihoodProfiler}
& Profile-likelihood analysis \\

and uncertainty
& \jlcmd{Turing.jl}~\cite{Ge2018Turing}
& Probabilistic programming \\

& \jlcmd{DiffEqBayes.jl}~\cite{DiffEqBayes}
& Bayesian ODE inference \\

Sensitivity analysis
& \jlcmd{GlobalSensitivity.jl}~\cite{Dixit2022GlobalSensitivity}
& Global sensitivity analysis \\

& \jlcmd{SciMLSensitivity.jl}~\cite{rackauckas2020universal}
& Forward and adjoint sensitivities \\

\bottomrule
\end{tabularx}
\end{table}

The workflow described above suggests three practical recommendations. First, perform structural identifiability analysis before sensitivity analysis or parameter estimation. Attempting to estimate structurally non-identifiable parameters cannot resolve the underlying ambiguity, regardless of the amount or quality of the data. Second, when non-identifiability is detected, use \jlcmd{find_identifiable_functions} to obtain an identifiable reparameterization before proceeding with parameter estimation or Bayesian inference (Section~\ref{sec:reparam}). Finally, treat structural identifiability, sensitivity analysis, and practical identifiability as complementary tools. Structural identifiability establishes whether parameters are theoretically recoverable, sensitivity analysis quantifies their influence on the model outputs, and practical identifiability evaluates how well they can be estimated from finite, noisy data.

\section{Reproducibility and Open Science}\label{sec:reproducibility}

\subsection{Reproducible computational research}

Reproducibility is a fundamental principle of computational science~\cite{Sandve2013}. Structural identifiability results should therefore be reported with sufficient information to allow independent verification. Reproducing an identifiability analysis requires not only the model equations and measurement scheme, but also the software environment used to perform the analysis.

\subsection{Open-source software}

The Julia language~\cite{Bezanson2017}, \jlcmd{StructuralIdentifiability.jl}, and the broader SciML ecosystem~\cite{Rackauckas2017} are open-source software projects. Their availability promotes transparency, accessibility, and reproducibility while removing the licensing restrictions associated with proprietary symbolic computation software.

\subsection{Scripts, notebook, and computational environment}\label{subsec:environment}

The companion material consists of one Julia script for each example in this tutorial, a Jupyter notebook that presents the case studies interactively, the SBML file of Section~\ref{subsec:mtk}, two figure scripts, and the files \texttt{Project.toml} and \texttt{Manifest.toml}, which record the versions of all packages. Table~\ref{tab:scripts} relates the scripts to the sections of this paper. From the root directory of the package, the commands
\begin{lstlisting}
julia --project=. -e "using Pkg; Pkg.instantiate()"   # install the recorded versions
julia --project=. run_all.jl                          # run every script
julia --project=. figures/make_figures.jl             # regenerate the data figures
\end{lstlisting}
install the environment, run all scripts while writing their output to the directory \texttt{expected\_output}, and regenerate Figures~\ref{fig:verdicts} and~\ref{fig:nonidentifiability}. The distributed output files allow a new run to be compared with ours line by line. Each script seeds Julia's random number generator, and \jlcmd{find_identifiable_functions} seeds its own (Section~\ref{subsec:api}), so a rerun in the recorded environment reproduces the distributed output apart from the measured runtimes. The listing in Section~\ref{subsec:viral} that uses \jlcmd{with_states=true} was added after version~2.1.0 of the package was archived; it runs unchanged in the same environment, and the script that produced it is supplied with this submission.

\begin{table}[!ht]
\centering
\caption{Companion scripts and the sections of this paper they reproduce.}
\label{tab:scripts}
\renewcommand{\arraystretch}{1.15}
\setlength{\tabcolsep}{5pt}
\small
\begin{tabularx}{\linewidth}{@{}l X l@{}}
\toprule
\rowcolor{tabhead}
\textbf{Script} & \textbf{Content} & \textbf{Section} \\
\midrule
\texttt{01\_exponential\_decay.jl} & Exponential decay & \ref{subsec:expdecay} \\
\texttt{02\_sir.jl} & SIR model & \ref{subsec:sir} \\
\texttt{03\_two\_compartment\_pk.jl} & Two-compartment model, three measurement schemes & \ref{subsec:pk} \\
\texttt{04\_viral\_dynamics.jl} & Viral dynamics, second output, known $T(0)$ & \ref{subsec:viral} \\
\texttt{05\_siwr.jl} & SIWR model, four measurement schemes & \ref{subsec:siwr} \\
\texttt{06\_seirh.jl} & SEIR-H model, fixed $\sigma$, occupancy & \ref{subsec:seirh} \\
\texttt{07\_bilinear.jl} & Bilinear model & \ref{sec:reparam} \\
\texttt{08\_modelingtoolkit\_catalyst.jl} & ModelingToolkit and Catalyst routes & \ref{subsec:mtk} \\
\texttt{09\_sbml\_import.jl} & SBML import of an enzyme mechanism & \ref{subsec:mtk} \\
\texttt{10\_multiple\_conditions.jl} & Pooling two experimental conditions & \ref{subsec:limitations} \\
\texttt{11\_rationalization.jl} & Two liftings of a non-rational model & \ref{subsec:limitations} \\
\texttt{12\_timing.jl} & Runtimes of the local and global tests & \ref{sec:best} \\
\texttt{13\_io\_coefficients.jl} & Example~2.14 of Hong \emph{et al.} & \ref{subsec:io} \\
\texttt{14\_siwr\_exchange.jl} & Simulation check of the SIWR symmetry & \ref{subsec:siwr} \\
\texttt{figures/make\_figures.jl} & Figures~\ref{fig:verdicts} and~\ref{fig:nonidentifiability} & \ref{subsec:consolidated}, \ref{sec:reparam} \\
\bottomrule
\end{tabularx}
\end{table}

The environment was rebuilt and tested for this revision, and the process illustrates why a manifest file is needed. The environment of the first revision fixed Julia~1.12.2, \jlcmd{StructuralIdentifiability.jl} 0.5.26, \jlcmd{ModelingToolkit.jl} 11.38.1, \jlcmd{OrdinaryDiffEq.jl} 7.3.0, and \jlcmd{Plots.jl} 1.41.6, but its manifest did not contain \jlcmd{Catalyst.jl}, and it could not be extended with new packages without resolving the dependencies again. After resolution, \jlcmd{StructuralIdentifiability.jl} 0.5.26 failed in its Wronskian computation, because the dependency \jlcmd{TimerOutputs.jl}, whose declared compatibility range admits versions 0.5 and 1, had been updated to version 1.2.2. Restricting \jlcmd{TimerOutputs.jl} to version 0.5.29 restored correct operation. The tested environment of this revision therefore keeps the package versions listed above, adds \jlcmd{Catalyst.jl} 16.4.0, \jlcmd{SBMLImporter.jl} 4.1.2, and \jlcmd{ReactionNetworkImporters.jl} 1.5.0, and records the restriction on \jlcmd{TimerOutputs.jl} in \texttt{Project.toml}. All scripts were run in this environment under Windows~11, and every case-study verdict agrees with the outputs of the first revision. A user who installs the packages without the manifest may obtain other dependency versions and, as this episode shows, possibly a non-working installation.

\subsection{Reproducibility checklist}

To facilitate reproducibility, structural identifiability studies should report the software environment together with the mathematical model. Table~\ref{tab:reproducibility} summarizes the recommended information.

\begin{table}[!ht]
\centering
\caption{Recommended information for reproducing structural identifiability analyses.}
\label{tab:reproducibility}
\renewcommand{\arraystretch}{1.25}
\setlength{\tabcolsep}{5pt}
\small
\begin{tabularx}{\linewidth}{@{}l X@{}}
\toprule
\rowcolor{tabhead}
\textbf{Item} & \textbf{Purpose} \\
\midrule
Julia version & Ensures reproducibility of the computational environment. \\
\jlcmd{StructuralIdentifiability.jl} version & Specifies the algorithms and implementation used. \\
\texttt{Project.toml} & Records direct package dependencies. \\
\texttt{Manifest.toml} & Pins exact versions of all dependencies. \\
Operating system & Documents the execution environment. \\
Git commit hash & Identifies the exact analysis code. \\
Zenodo DOI & Provides a permanent archive of the computational resources. \\
Random seed & Allows exact reproduction of probabilistic analyses. \\
\bottomrule
\end{tabularx}
\end{table}

\subsection{Data and code availability}

The complete reproducibility package is archived on Zenodo~\cite{Alsammani2026} as version~2.1.0, \url{https://doi.org/10.5281/zenodo.23093726}, and is developed in a public GitHub repository. This version contains the scripts of Table~\ref{tab:scripts} with their expected output, the notebook, the SBML model file, the figure scripts, \texttt{Project.toml}, \texttt{Manifest.toml}, \texttt{README.md}, and \texttt{LICENSE}.

\subsection{Educational value}

Reproducible notebooks provide an effective teaching resource. Because model equations, measurement schemes, and identifiability results can be modified interactively, students can immediately observe how changes in model structure or experimental design affect structural identifiability. This interactive workflow is particularly valuable for graduate instruction in mathematical biology, epidemiology, and systems biology.
\section{Discussion}\label{sec:discussion}

Structural identifiability is a fundamental prerequisite for mechanistic modeling because it determines whether model parameters can, in principle, be uniquely recovered from the available observations. Although often regarded as a theoretical property of differential equation models, it has direct practical consequences for parameter estimation, model interpretation, uncertainty quantification, and experimental design. The examples presented throughout this tutorial demonstrate that structural non-identifiability is common across epidemiology, pharmacokinetics, and within-host viral dynamics, even for relatively simple models. More importantly, the results show that identifiability depends jointly on the mathematical model and the measurement scheme. A model that is non-identifiable under one observation protocol may become globally identifiable after additional measurements, modified experimental conditions, or an appropriate reparameterization.

The tutorial presents a unified and reproducible workflow for symbolic structural identifiability analysis using \jlcmd{StructuralIdentifiability.jl}, and pairs every software verdict with an elementary explanation. By integrating model construction, symbolic analysis, numerical simulation, and reproducible computational practices within the Julia SciML ecosystem, the workflow lowers the practical barriers to incorporating identifiability analysis into routine modeling studies. The case studies illustrate that symbolic analysis should not be viewed solely as a diagnostic tool but also as a guide for model development. Identifiable parameter combinations suggest meaningful reparameterizations, while non-identifiable parameters indicate where additional measurements or revised experimental designs are required. Consequently, structural identifiability analysis provides actionable information before computationally intensive parameter estimation, Bayesian inference, or uncertainty quantification is undertaken.

The tutorial also highlights several broader lessons. First, measurement design is often more important than model complexity for achieving parameter identifiability. Additional observables, informative experimental conditions, or independent biological measurements can substantially improve parameter recoverability without altering the underlying model. Second, identifiable parameter combinations frequently represent the appropriate quantities for estimation and interpretation when individual parameters cannot be uniquely determined. Finally, structural identifiability should be regarded as the first stage of a complete inference pipeline. Once structural identifiability has been established, complementary methods such as profile likelihood analysis, Bayesian inference, and uncertainty quantification can be used to assess practical parameter recoverability from finite and noisy data~\cite{Raue2009,Wieland2021,Hines2014}.

Several limitations remain. The framework considered here is restricted to deterministic rational ODE models and therefore does not directly address stochastic differential equations, delay differential equations, partial differential equations, or hybrid mechanistic--machine learning models, although some hybrid models can be rewritten exactly as rational models (Section~\ref{subsec:limitations}). In addition, structural identifiability alone cannot assess practical identifiability, model misspecification, or the adequacy of the underlying biological assumptions. Current symbolic methods also become increasingly demanding for large, densely connected nonlinear systems and offer no dedicated interface for pooling experiments performed under different conditions, although such designs can be analyzed by augmenting the model (Section~\ref{subsec:limitations}). These limitations identify important directions for future methodological development rather than diminishing the value of structural identifiability analysis.

Overall, advances in symbolic algorithms, open-source software, and the Julia scientific computing ecosystem have transformed structural identifiability analysis from a specialized theoretical subject into a practical component of modern computational modeling. Incorporating structural identifiability analysis early in the modeling workflow can improve model formulation, guide experimental design, reduce unnecessary parameter estimation, and increase the reliability and reproducibility of mechanistic modeling studies across the biological and biomedical sciences.
\section{Conclusion}\label{sec:conclusion}

This tutorial presented a reproducible computational framework for symbolic structural identifiability analysis of mechanistic ordinary differential equation models using \jlcmd{StructuralIdentifiability.jl}. By combining mathematical foundations with practical case studies, it demonstrated how symbolic identifiability analysis can be integrated into a modern modeling workflow, from model formulation and structural analysis to simulation, parameter estimation, and uncertainty quantification within the Julia SciML ecosystem. The examples illustrate globally and locally identifiable models, structural non-identifiability, identifiable parameter combinations, and practical strategies for restoring identifiability through additional measurements and model reparameterization.

Beyond introducing computational tools, this tutorial emphasizes a broader principle: structural identifiability should be regarded as a fundamental component of mechanistic model development rather than a post hoc diagnostic. Performing identifiability analysis before parameter estimation improves model formulation, informs measurement and experimental design, reduces unnecessary computational effort, and supports more reliable scientific inference.

The continued development of symbolic algorithms and open-source software is expanding the scope and accessibility of structural identifiability analysis across mathematical biology, epidemiology, pharmacokinetics, systems biology, and related fields. We hope this tutorial provides both a practical reference and a foundation for researchers seeking to incorporate structural identifiability analysis into reproducible computational modeling workflows.
\section*{Acknowledgments}

The author gratefully acknowledges the developers of \jlcmd{StructuralIdentifiability.jl} and the broader Julia SciML community for developing and maintaining the open-source software that made this work possible. The author also sincerely thanks the Editor and the anonymous reviewers for their careful evaluation and constructive comments, which substantially improved the clarity, accuracy, and scope of this tutorial.

\section*{Data and Code Availability}

The complete reproducibility package is publicly available under the MIT License, permanently archived on Zenodo~\cite{Alsammani2026} (version~2.1.0), and maintained at \url{https://github.com/aalsammani/Structural_Identifiability_Tutorial_Julia}. It includes all Julia scripts and expected outputs, the tutorial notebook, the SBML model, the figure-generation scripts, and the \texttt{Project.toml} and \texttt{Manifest.toml} files defining the tested computational environment described in Section~\ref{subsec:environment}.



\begin{thebibliography}{99}
	
	\bibitem{Raue2009}
	A.~Raue, C.~Kreutz, T.~Maiwald, J.~Bachmann, M.~Schilling,
	U.~Klingm\"uller, and J.~Timmer,
	``Structural and practical identifiability analysis of partially observed
	dynamical models by exploiting the profile likelihood,''
	\emph{Bioinformatics}, vol.~25, no.~15, pp.~1923--1929, 2009.
	doi: \url{https://doi.org/10.1093/bioinformatics/btp358}.
	
	\bibitem{Wieland2021}
	F.-G.~Wieland, A.~L.~Hauber, M.~Rosenblatt, C.~T\"onsing, and J.~Timmer,
	``On structural and practical identifiability,''
	\emph{Current Opinion in Systems Biology}, vol.~25, pp.~60--69, 2021.
	doi: \url{https://doi.org/10.1016/j.coisb.2021.03.005}.

    \bibitem{Alsammani2026}
    A.~Alsammani,
    ``Structural Identifiability Tutorial in Julia,'' version 2.1.0 (reproducibility package for this tutorial),
    \emph{Zenodo}, 2026.
    doi: \url{https://doi.org/10.5281/zenodo.23093726}.

	\bibitem{Tuncer2018}
	N.~Tuncer and T.~T.~Le,
	``Structural and practical identifiability analysis of outbreak models,''
	\emph{Mathematical Biosciences}, vol.~299, pp.~1--18, 2018.
	doi: \url{https://doi.org/10.1016/j.mbs.2018.02.004}.
	
	\bibitem{BellmanAstrom1970}
	R.~Bellman and K.~J.~\AA str\"om,
	``On structural identifiability,''
	\emph{Mathematical Biosciences}, vol.~7, no.~3--4,
	pp.~329--339, 1970.
	doi: \url{https://doi.org/10.1016/0025-5564(70)90132-X}.

\bibitem{Alsammani2023HBV}
	A.~Alsammani,
	``Mathematical analysis of autonomous and nonautonomous hepatitis B virus transmission models,''
	in \emph{Computational Science and Its Applications -- ICCSA 2023},
	Lecture Notes in Computer Science, vol.~14105,
	pp.~327--343, 2023.
	doi: \url{https://link.springer.com/chapter/10.1007/978-3-031-37108-0_21}.

\bibitem{Alsammani2025Vaccination}
	A.~Alsammani, C.~N.~Ngonghala, and M.~Martcheva,
	``Impact of vaccination behavior on COVID-19 dynamics and economic outcomes,''
	\emph{Mathematical Biosciences and Engineering}, vol.~22, no.~9,
	pp.~2300--2338, 2025.
	doi: \url{https://doi.org/10.3934/mbe.2025084}.
	
	\bibitem{Pohjanpalo1978}
	H.~Pohjanpalo,
	``System identifiability based on the power series expansion of the
	solution,''
	\emph{Mathematical Biosciences}, vol.~41, no.~1--2,
	pp.~21--33, 1978.
	doi: \url{https://doi.org/10.1016/0025-5564(78)90063-9}.
	
	\bibitem{Cobelli1980}
	C.~Cobelli and J.~J.~DiStefano~III,
	``Parameter and structural identifiability concepts and ambiguities:
	A critical review and analysis,''
	\emph{American Journal of Physiology-Regulatory, Integrative and
		Comparative Physiology}, vol.~239, no.~1,
	pp.~R7--R24, 1980.
	doi: \url{https://doi.org/10.1152/ajpregu.1980.239.1.R7}.
	
	\bibitem{Walter1997}
	E.~Walter and L.~Pronzato,
	\emph{Identification of Parametric Models from Experimental Data}.
	Springer, 1997.
	
	\bibitem{Ljung1994}
	L.~Ljung and T.~Glad,
	``On global identifiability for arbitrary model parametrizations,''
	\emph{Automatica}, vol.~30, no.~2,
	pp.~265--276, 1994.
	doi: \url{https://doi.org/10.1016/0005-1098(94)90029-9}.
	
	\bibitem{Miao2011}
	H.~Miao, X.~Xia, A.~S.~Perelson, and H.~Wu,
	``On identifiability of nonlinear ODE models and applications in viral
	dynamics,''
	\emph{SIAM Review}, vol.~53, no.~1,
	pp.~3--39, 2011.
	doi: \url{https://doi.org/10.1137/090757009}.
	
	\bibitem{Bellu2007}
	G.~Bellu, M.~P.~Saccomani, S.~Audoly, and L.~D'Angi\`o,
	``DAISY: A new software tool to test global identifiability of biological
	and physiological systems,''
	\emph{Computer Methods and Programs in Biomedicine},
	vol.~88, no.~1, pp.~52--61, 2007.
	doi: \url{https://doi.org/10.1016/j.cmpb.2007.07.002}.
	
	\bibitem{Villaverde2016}
	A.~F.~Villaverde, A.~Barreiro, and A.~Papachristodoulou,
	``Structural identifiability of dynamic systems biology models,''
	\emph{PLoS Computational Biology},
	vol.~12, no.~10, p.~e1005153, 2016.
	doi: \url{https://doi.org/10.1371/journal.pcbi.1005153}.
	
	\bibitem{Hong2020}
	H.~Hong, A.~Ovchinnikov, G.~Pogudin, and C.~Yap,
	``Global identifiability of differential models,''
	\emph{Communications on Pure and Applied Mathematics},
	vol.~73, no.~9, pp.~1831--1879, 2020.
	doi: \url{https://doi.org/10.1002/cpa.21921}.
	
	\bibitem{Dong2023}
	R.~Dong, C.~Goodbrake, H.~A.~Harrington, and G.~Pogudin,
	``Differential elimination for dynamical models via projections with
	applications to structural identifiability,''
	\emph{SIAM Journal on Applied Algebra and Geometry},
	vol.~7, no.~1, pp.~194--235, 2023.
	doi: \url{https://doi.org/10.1137/22M1469067}.
	
	\bibitem{Ligon2018}
	T.~S.~Ligon, F.~Fr\"ohlich, O.~T.~Chi\c{s}, J.~R.~Banga,
	E.~Balsa-Canto, and J.~Hasenauer,
	``GenSSI~2.0: Multi-experiment structural identifiability analysis of
	SBML models,''
	\emph{Bioinformatics},
	vol.~34, no.~8, pp.~1421--1423, 2018.
	doi: \url{https://doi.org/10.1093/bioinformatics/btx735}.
	
	\bibitem{Meshkat2009}
	N.~Meshkat, M.~Eisenberg, and J.~J.~DiStefano~III,
	``An algorithm for finding globally identifiable parameter combinations
	of nonlinear ODE models using Gr\"obner bases,''
	\emph{Mathematical Biosciences},
	vol.~222, no.~2, pp.~61--72, 2009.
	doi: \url{https://doi.org/10.1016/j.mbs.2009.08.010}.

	\bibitem{Meshkat2014}
	N.~Meshkat, C.~E.~Kuo, and J.~DiStefano~III,
	``On finding and using identifiable parameter combinations in nonlinear
	dynamic systems biology models and COMBOS: A novel web implementation,''
	\emph{PLoS ONE}, vol.~9, no.~10, p.~e110261, 2014.
	doi: \url{https://doi.org/10.1371/journal.pone.0110261}.
	
	\bibitem{Hong2019SIAN}
	H.~Hong, A.~Ovchinnikov, G.~Pogudin, and C.~Yap,
	``SIAN: Software for structural identifiability analysis of ODE models,''
	\emph{Bioinformatics},
	vol.~35, no.~16, pp.~2873--2874, 2019.
	doi: \url{https://doi.org/10.1093/bioinformatics/bty1069}.
	
	\bibitem{Chis2011}
	O.-T.~Chi\c{s}, J.~R.~Banga, and E.~Balsa-Canto,
	``Structural identifiability of systems biology models:
	A critical comparison of methods,''
	\emph{PLoS ONE},
	vol.~6, no.~11, p.~e27755, 2011.
	doi: \url{https://doi.org/10.1371/journal.pone.0027755}.
	
	\bibitem{Ovchinnikov2022}
	A.~Ovchinnikov, A.~Pillay, G.~Pogudin, and T.~Scanlon,
	``Computing all identifiable functions of parameters for ODE models,''
	\emph{Systems \& Control Letters},
	vol.~157, p.~105030, 2021.
	doi: \url{https://doi.org/10.1016/j.sysconle.2021.105030}.
	
	\bibitem{DeminPogudin2026}
	A.~Demin and G.~Pogudin,
	``Simple generators of rational function fields,''
	\emph{arXiv preprint}, arXiv:2602.10878, 2026.

	\bibitem{DeminPogudin2025}
	A.~Demin and G.~Pogudin,
	``Finding simple generators of rational function fields,''
	\emph{ACM Communications in Computer Algebra},
	vol.~59, no.~4, pp.~138--146, 2025.
	doi: \url{https://doi.org/10.1145/3832046.3832062}.

	\bibitem{DeminPogudinRackauckas2026}
	A.~Demin, G.~Pogudin, and C.~Rackauckas,
	``Observable functions of rational ODE models and how to find them,''
	\emph{arXiv preprint}, arXiv:2609.06134, 2026.
	doi: \url{https://doi.org/10.48550/arXiv.2609.06134}.

	\bibitem{HermannKrener1977}
	R.~Hermann and A.~J.~Krener,
	``Nonlinear controllability and observability,''
	\emph{IEEE Transactions on Automatic Control},
	vol.~22, no.~5, pp.~728--740, 1977.
	doi: \url{https://doi.org/10.1109/TAC.1977.1101601}.
	
	\bibitem{OvchinnikovPogudinThompson2021}
	A.~Ovchinnikov, G.~Pogudin, and P.~Thompson,
	``Parameter identifiability and input--output equations,''
	\emph{Applicable Algebra in Engineering, Communication and Computing},
	vol.~34, no.~2, pp.~165--182, 2023 (published online 2021).
	doi: \url{https://doi.org/10.1007/s00200-021-00486-8}.
	
	\bibitem{Rackauckas2017}
	C.~Rackauckas and Q.~Nie,
	``DifferentialEquations.jl: A performant and feature-rich ecosystem for
	solving differential equations in Julia,''
	\emph{Journal of Open Research Software},
	vol.~5, no.~1, p.~15, 2017.
	doi: \url{https://doi.org/10.5334/jors.151}.
	
	\bibitem{Bezanson2017}
	J.~Bezanson, A.~Edelman, S.~Karpinski, and V.~B.~Shah,
	``Julia: A fresh approach to numerical computing,''
	\emph{SIAM Review},
	vol.~59, no.~1, pp.~65--98, 2017.
	doi: \url{https://doi.org/10.1137/141000671}.
	
	\bibitem{Sedoglavic2002}
	A.~Sedoglavic,
	``A probabilistic algorithm to test local algebraic observability in
	polynomial time,''
	\emph{Journal of Symbolic Computation},
	vol.~33, no.~5, pp.~735--755, 2002.
	doi: \url{https://doi.org/10.1006/jsco.2002.0532}.
	
    \bibitem{Liyanage2025}
    Y.~R.~Liyanage, O.~Saucedo, N.~Tuncer, and G.~Chowell,
    ``A tutorial on structural identifiability of epidemic models using
    \jlcmd{StructuralIdentifiability.jl},''
    \emph{Bulletin of Mathematical Biology}, vol.~88, no.~6,
    Art.~102, 2026.
    doi: \url{https://doi.org/10.1007/s11538-026-01648-w}.
	
	\bibitem{Kermack1927}
	W.~O.~Kermack and A.~G.~McKendrick,
	``A contribution to the mathematical theory of epidemics,''
	\emph{Proceedings of the Royal Society A},
	vol.~115, no.~772, pp.~700--721, 1927.
	doi: \url{https://doi.org/10.1098/rspa.1927.0118}.
	
	\bibitem{Perelson1996}
	A.~S.~Perelson, A.~U.~Neumann, M.~Markowitz,
	J.~M.~Leonard, and D.~D.~Ho,
	``HIV-1 dynamics in vivo: Virion clearance rate, infected cell life span,
	and viral generation time,''
	\emph{Science},
	vol.~271, no.~5255, pp.~1582--1586, 1996.
	doi: \url{https://doi.org/10.1126/science.271.5255.1582}.
	
	\bibitem{Eisenberg2013}
	M.~C.~Eisenberg, S.~L.~Robertson, and J.~H.~Tien,
	``Identifiability and estimation of multiple transmission pathways in
	cholera and waterborne disease,''
	\emph{Journal of Theoretical Biology},
	vol.~324, pp.~84--102, 2013.
	doi: \url{https://doi.org/10.1016/j.jtbi.2012.12.021}.
	
	\bibitem{Villaverde2019}
	A.~F.~Villaverde,
	``Observability and structural identifiability of nonlinear biological systems,''
	\emph{Complexity}, vol.~2019, Art.~8497093, 2019.
	doi: \url{https://doi.org/10.1155/2019/8497093}.
	
	\bibitem{Hines2014}
	K.~E.~Hines, T.~R.~Middendorf, and R.~W.~Aldrich,
	``Determination of parameter identifiability in nonlinear biophysical models:
	A Bayesian approach,''
	\emph{The Journal of General Physiology},
	vol.~143, no.~3, pp.~401--416, 2014.
	doi: \url{https://doi.org/10.1085/jgp.201311116}.
	
	\bibitem{Eisenberg2014}
	M.~C.~Eisenberg and M.~A.~L.~Hayashi,
	``Determining identifiable parameter combinations using subset profiling,''
	\emph{Mathematical Biosciences},
	vol.~256, pp.~116--126, 2014.
	doi: \url{https://doi.org/10.1016/j.mbs.2014.08.008}.
	
	\bibitem{Chen2018}
	R.~T.~Q.~Chen, Y.~Rubanova, J.~Bettencourt, and D.~Duvenaud,
	``Neural ordinary differential equations,''
	in \emph{Advances in Neural Information Processing Systems},
	vol.~31, 2018.
	doi: \url{https://doi.org/10.48550/arXiv.1806.07366}.
	
	\bibitem{Sandve2013}
	G.~K.~Sandve, A.~Nekrutenko, J.~Taylor, and E.~Hovig,
	``Ten simple rules for reproducible computational research,''
	\emph{PLoS Computational Biology},
	vol.~9, no.~10, p.~e1003285, 2013.
	doi: \url{https://doi.org/10.1371/journal.pcbi.1003285}.

	\bibitem{Daneker2023SBINN}
	M.~Daneker, Z.~Zhang, G.~E.~Karniadakis, and L.~Lu,
	``Systems biology: Identifiability analysis and parameter identification
	via systems-biology-informed neural networks,''
	in \emph{Computational Modeling of Signaling Networks}, Methods in
	Molecular Biology, Springer US, 2023, pp.~87--105.
	doi: \url{https://doi.org/10.1007/978-1-0716-3008-2_4}.

	\bibitem{Norden2025}
	J.~Norden, E.~Oostwal, M.~Chappell, P.~Tino, and K.~Bunte,
	``Structure is information: Structural identifiability mappings for
	machine learning with partially observed dynamical systems,''
	\emph{IEEE Transactions on Cybernetics}, vol.~56, no.~10, pp.~5423--5435, 2026.
	doi: \url{https://doi.org/10.1109/TCYB.2026.3694145}; preprint arXiv:2502.04131.

	\bibitem{Qian2024CoaguloNet}
	Y.~Qian, G.~Zhu, Z.~Zhang, S.~Modepalli, Y.~Zheng, X.~Zheng, G.~Frydman,
	and H.~Li,
	``Coagulo-Net: Enhancing the mathematical modeling of blood coagulation
	using physics-informed neural networks,''
	\emph{Neural Networks}, vol.~180, p.~106732, 2024.
	doi: \url{https://doi.org/10.1016/j.neunet.2024.106732}.

	\bibitem{Ma2021MTK}
	Y.~Ma, S.~Gowda, R.~Anantharaman, C.~Laughman, V.~Shah, and C.~Rackauckas,
	``ModelingToolkit: A composable graph transformation system for
	equation-based modeling,''
	\emph{arXiv preprint}, arXiv:2103.05244, 2021.

	\bibitem{Loman2023Catalyst}
	T.~E.~Loman, Y.~Ma, V.~Ilin, S.~Gowda, N.~Korsbo, N.~Yewale,
	C.~Rackauckas, and S.~A.~Isaacson,
	``Catalyst: Fast and flexible modeling of reaction networks,''
	\emph{PLoS Computational Biology}, vol.~19, no.~10, p.~e1011530, 2023.
	doi: \url{https://doi.org/10.1371/journal.pcbi.1011530}.

	\bibitem{Bostrom2024SBMLImporter}
	\jlcmd{SBMLImporter.jl}: a Julia package that imports SBML models as
	\jlcmd{Catalyst.jl} \jlcmd{ReactionSystem} objects, version~4.1.2;
	software and documentation at \url{https://github.com/sebapersson/SBMLImporter.jl}.
	The package documentation asks users to cite~\cite{Persson2024PEtab}.

	\bibitem{ReactionNetworkImporters}
	\jlcmd{ReactionNetworkImporters.jl}: a Julia package that imports
	reaction networks specified in the BioNetGen \texttt{.net} format or as
	stoichiometric/complex matrices into \jlcmd{Catalyst.jl}
	\jlcmd{ReactionSystem} objects, developed by the SciML organization and
	documented at \url{https://github.com/SciML/ReactionNetworkImporters.jl}.

	\bibitem{Persson2024PEtab}
	S.~Persson, F.~Fr\"ohlich, S.~Grein, T.~Loman, D.~Ognissanti,
	V.~Hasselgren, J.~Hasenauer, and M.~Cvijovic,
	``PEtab.jl: advancing the efficiency and utility of dynamic
	modelling,''
	\emph{Bioinformatics}, vol.~41, no.~9, p.~btaf497, 2025.
	doi: \url{https://doi.org/10.1093/bioinformatics/btaf497}.

	\bibitem{Dixit2022GlobalSensitivity}
	V.~K.~Dixit and C.~Rackauckas,
	``GlobalSensitivity.jl: Performant and parallel global sensitivity
	analysis with Julia,''
	\emph{Journal of Open Source Software}, vol.~7, no.~76, p.~4561, 2022.
	doi: \url{https://doi.org/10.21105/joss.04561}.


	\bibitem{DiffEqBayes}
	\jlcmd{DiffEqBayes.jl}: extension functionality that uses
	\jlcmd{Turing.jl}, \texttt{Stan.jl}, \texttt{DynamicHMC.jl}, and
	\texttt{ApproxBayes.jl} to perform Bayesian estimation of
	\jlcmd{DifferentialEquations.jl} models, developed by the SciML
	organization and documented at
	\url{https://github.com/SciML/DiffEqBayes.jl}.

	\bibitem{Ge2018Turing}
	H.~Ge, K.~Xu, and Z.~Ghahramani,
	``Turing: A language for flexible probabilistic inference,''
	in \emph{International Conference on Artificial Intelligence and
	Statistics (AISTATS)}, Proceedings of Machine Learning Research,
	vol.~84, pp.~1682--1690, 2018.

	\bibitem{Borisov2020LikelihoodProfiler}
	I.~Borisov and E.~Metelkin,
	``Confidence intervals by constrained optimization---An algorithm and
	software package for practical identifiability analysis in systems
	biology,''
	\emph{PLoS Computational Biology}, vol.~16, no.~12, p.~e1008495, 2020.
	doi: \url{https://doi.org/10.1371/journal.pcbi.1008495};
	implemented as \jlcmd{LikelihoodProfiler.jl},
	\url{https://github.com/insysbio/LikelihoodProfiler.jl}.

	\bibitem{LiyanageViruses2025}
	Y.~R.~Liyanage, G.~Chowell, G.~Pogudin, and N.~Tuncer,
	``Structural and practical identifiability of phenomenological growth
	models for epidemic forecasting,''
	\emph{Viruses}, vol.~17, no.~4, p.~496, 2025.
	doi: \url{https://doi.org/10.3390/v17040496}.

	\bibitem{Pant2026}
	B.~Pant, O.~Saucedo, and G.~Pogudin,
	``Uncovering identifiability of epidemiological models: basic
	reproduction number and complementary data streams,''
	\emph{medRxiv preprint}, 2026.
	doi: \url{https://doi.org/10.64898/2026.01.16.26344284}.

    \bibitem{dixit2023optimization}
    V.~K. Dixit and C. Rackauckas,
    ``Optimization.jl: A unified optimization package,''
    Zenodo, version 3.12.1, 2023.
    doi: \url{https://doi.org/10.5281/zenodo.7738525}.
    
    \bibitem{rackauckas2020universal}
    C. Rackauckas, Y. Ma, J. Martensen, C. Warner, K. Zubov, R. Supekar,
    D. Skinner, A. Ramadhan, and A. Edelman,
    ``Universal differential equations for scientific machine learning,''
    \textit{arXiv preprint arXiv:2001.04385}, 2020.
    doi: \url{https://doi.org/10.48550/arXiv.2001.04385}.

\end{thebibliography}
\end{document}